\documentclass[12pt,reprint,showpacs,groupedaddress,eqsecnum,amsfonts,amsmath,amssymb,aps,prd]{revtex4}
\usepackage{graphicx}
\usepackage{latexsym}
\usepackage{amssymb}
\usepackage{mathrsfs}
\usepackage{amsmath}
\usepackage{amsthm}
\usepackage{color}
\usepackage{dcolumn}
\usepackage{bm}

\usepackage[
text={7in,20in},centering,
margin=2in,
total={6.5in,8.75in},
top=1.2in,
left=1.1in,
right=1.1in,
includefoot,
a3paper,
hmargin={3cm,0.8in},
]{geometry}

\definecolor{dyellow}{rgb}{1.,0.8,.0}
\definecolor{myblue}{rgb}{.1,.1,.7}
\definecolor{dcyan}{rgb}{.0,.6,.6}
\definecolor{dmagenta}{rgb}{0.6,0.0,0.6}
\definecolor{brown}{rgb}{0.6,0.2,0.}
\definecolor{darkblue}{rgb}{.0,.0,0.5}
\definecolor{darkred}{rgb}{0.75,0.0,0.0}
\definecolor{orange}{rgb}{1.,.6,.0}
\definecolor{dorange}{rgb}{0.8,.4,.0}
\definecolor{darkgreen}{rgb}{0.0,0.6,0.0}
\definecolor{purple}{rgb}{.4,.0,.4}
\definecolor{grey}{rgb}{0.5,0.5,0.5}

\begin{document}
\hyphenpenalty=1000
\preprint{APS/123-QED}
\title{Analytical analysis on the orbits of Taiji spacecrafts}

\newcommand*{\PKU}{School of Physics, University of Chinese Academy of Sciences, Beijing 100049, China}\affiliation{\PKU}
\newcommand*{\INFN}{Institute of High Energy Physics,
Chinese Academy of Sciences, Beijing, 100049, China}\affiliation{\INFN}
\newcommand*{\CICQM}{CAS Center for Excellence in Particle Physics, Beijing 100049, China}\affiliation{\CICQM}
\newcommand*{\CHEP}{}\affiliation{\CHEP}

\author{Bofeng Wu}\email{bofengw@pku.edu.cn}\affiliation{\PKU}
\author{Chao-Guang Huang}\email{huangcg@ihep.ac.cn}\affiliation{\INFN,\PKU}
\author{Cong-Feng Qiao}\email{ qiaocf@ucas.ac.cn}\affiliation{\PKU,\CICQM}

\begin{abstract}
The unperturbed Keplerian orbits of Taiji spacecrafts are expanded to $e^3$ order in the heliocentric coordinate system, where $e$ is their orbital eccentricity. The three arm-lengths of Taiji triangle and their rates of change are also expanded to $e^3$ order, while the three vertex angles are expanded to $e^2$ order. These kinematic indicators of Taiji triangle are, further, minimized, respectively, by adjusting the tilt angle of Taiji plane relative to the ecliptic plane around $\pm\pi/3$, and thus, their corresponding optimized expressions are presented. Then, under the case that the nominal trailing angle of Taiji constellation following the Earth is set to be $\chi(\approx\pm\pi/9)$ from the viewpoint of the Sun, the influence of the Earth perturbation on three spacecrafts is calculated according to the equations of motion in the problem of three bodies, and the perturbative solutions of the leading order and the next leading order are derived. With the perturbative solutions, the leading-order corrections to the above kinematic indicators of Taiji triangle and the expression of the above trailing angle to the order of $e^3$ are provided.
\end{abstract}
\maketitle
\section{Introduction\label{Sec:first}}
The successful detection of gravitational waves (GWs) by the LIGO and Virgo collaborations~\cite{TheLIGOScientific:2016agk} opens up the era of GW astronomy, and promotes the study of General Relativity (GR) and astrophysics~\cite{TheLIGOScientific:2016src,TheLIGOScientific:2016htt,GBM:2017lvd}.
Because of the disturbance of the gravity gradient noise, LIGO and Virgo are sensitive to GWs above $10$ Hz~\cite{Danzmann:1997hm,Adhikari:2013kya}, and in the future, some new ground-based detectors could make possible
GW observations down to $0.1$ Hz~\cite{Harms:2013raa}.
Even so, for GWs from $0.1$ mHz to $1.0$ Hz, the space-based GW detector like LISA~\cite{Dhurandhar:2004rv,Nayak:2006zm} becomes the next interesting target for the further study of GW.

After LISA mission was put forward, Chinese scientists began to show their interests in the space-based GW detector~\cite{xuefei2011,Gong:2014mca}, and Taiji program~\cite{Hu:2017mde,Wu:2018clg} was set up by the Chinese Academy of Sciences. Taiji program, like LISA, is based on three identical spacecrafts (SCs) orbiting the Sun, and these SCs form a triangle of side about $3\times10^6$ km.
Similarly to LISA, Taiji program will use
coherent laser beams exchanged between SCs to observe low GWs covering the ranges from $0.1$ mHz to $1.0$ Hz, and it is complementary to the ground-based GW detection program in an essential way, which is similar to the observation for the electromagnetic waves in different wavebands~\cite{Dhurandhar:2008yu}.
Maybe LISA and Taiji will be in operation at the same time for a period, and their simultaneous operation will certainly prompt the GW detection significantly.

Analytical treatment of the motion of SCs is important, because it is crucial for thoroughly studying optical links and light propagation between SCs. By using the analytical method, many problems related to the motion of SCs will be
more transparent than by using numerical simulations. The analytical treatment is also the basis for further numerical simulations. The analytical analysis on the motion of LISA SCs has been made in literatures~\cite{Dhurandhar:2004rv,Nayak:2006zm,Dhurandhar:2008yu,Pucacco:2010mn}.
In Refs.~\cite{Dhurandhar:2004rv,Nayak:2006zm}, the unperturbed Keplerian orbits of SCs are expanded to $\alpha^2$ order in the Hill system or the Clohessy-Wiltshire (CW) system~\cite{Pucacco:2010mn}, where the parameter $\alpha$ is proportional to the orbital eccentricity of SCs to the first order, and further, the arm-length of LISA triangle, formed by LISA SCs, and its rate of change are also expanded to $\alpha^2$ order and are minimized, respectively, by adjusting the tilt angle of LISA plane, in which LISA triangle lies, relative to the ecliptic plane around $\pi/3$. In Refs.~\cite{Dhurandhar:2008yu,Pucacco:2010mn}, the Earth's orbit is assumed to be a circle in the ecliptic plane, the nominal trailing angle of LISA constellation following the Earth is set to be $\pi/9$ from the viewpoint of the Sun, and the influence of the Earth perturbation on SCs is dealt with by the linear perturbative approach in the CW system. Seeing the similarity between Taiji program and LISA mission, the model in Refs.~\cite{Dhurandhar:2004rv,Nayak:2006zm,Dhurandhar:2008yu,Pucacco:2010mn} designed for LISA mission can be used in the Taiji program with the replacement of the orbit parameters if the higher precision is not needed.

However, the results to $\alpha^2$ order is still not enough.  The choice of the circular orbit means that the contribution of the eccentricity $e'$ of the Earth's orbit in the perturbation on LISA SCs is ignored.
Besides, due to the simplistic model, the obtained perturbative solution does not include the contribution of
the interaction between the Sun and the Earth either.  In this paper, we will analytically analyze the orbits of Taiji SCs in the heliocentric coordinate system in a higher precision than those of LISA SCs in the above related references.

In order to facilitate follow-up study in a higher precision, the unperturbed Keplerian orbits of Taiji SCs should be first studied in the heliocentric coordinate system as the primary task, and these orbits are expanded to the cube of eccentricity $e$. Then the three arm-lengths of  Taiji triangle, formed by SCs, and their rates of change are also expanded to $e^3$ order, while the three vertex angles of the triangle are expanded to $e^2$ order.  The expansions of these kinematic indicators of Taiji triangle show that its shape depends on the tilt angle of Taiji plane, in which Taiji triangle lies, with respect to the ecliptic plane and that all the kinematic indicators of Taiji triangle vary periodically over time. The main features of the unperturbed orbits can be summarized as follows.
\begin{itemize}
\item[$\bullet$] Up to $e^0$ order, the tilt angle remains constant angle $\phi$;
\item[$\bullet$] Under the cases of $\phi=\pm\pi/3$, Taiji triangle is approximately equilateral one, i.e.
\begin{itemize}
\item the three arm-lengths remain $2\sqrt{3}Re$ up to $e^1$ order, where $R$ is the semi-major axis of the elliptical orbits of SCs, and its value is equal to the semi-major axis of the Earth's orbit;
\item the three change rates of arm-lengths remain zero up to $e^1$ order;
\item the three vertex angles remain $\pi/3$ up to $e^0$ order.
\end{itemize}
\end{itemize}
For Taiji, the nominal arm-length is $3\times10^6$ km, and by the above conclusion, $e\approx5.789\times10^{-3}$.

As LISA~\cite{Dhurandhar:2008yu}, the laser frequency noise of Taiji is suppressed by time-delay interferometry (TDI).
The first generation TDI, however, works only for the stationary configuration in a flat spacetime. The relative motion between SCs may require modified first generation TDI or further, the second generation TDI~\cite{Dhurandhar:2008yu,Tinto:2003vj,Vallisneri:2005ji,Tinto:2014lxa}. As indicated by Ref.~\cite{Dhurandhar:2008yu}, the reasonably optimized model of LISA may be no need to use the second generation TDI so as to avoid possible difficulty caused by non-commuting time-delay operators in the data analysis.  Because the orbital eccentricity of Taiji SCs is smaller than that of LISA SCs (presented in Refs.~\cite{Dhurandhar:2004rv,Nayak:2006zm,Dhurandhar:2008yu}), if the model of Taiji is optimized, and namely, the amplitude of the relative motion between Taiji SCs is reduced, a simpler TDI strategy could be considered for Taiji program. Moreover, the relative motion between SCs will also cause the Doppler shift of the laser frequency, which will interfere the measurement of GWs~\cite{Nayak:2006zm}, and in view of this adverse effect, the orbits of SCs need to be
optimized to reduce the amplitude of the relative motion between SCs.
It can be shown that by adjusting the angle $\phi$ around $\pm\pi/3$ at $e^1$ order, namely,
\begin{equation}\label{equ1.1}
\phi=\pm\left(\frac{\pi}{3}+\frac{5\sqrt{3}}{8}e\right),
\end{equation}
the variations of all the kinematic indicators of Taiji triangle (the three arm-lengths and their rates of change,  the three vertex angles) can be minimized, respectively, which is compatible to that of LISA in Refs.~\cite{Nayak:2006zm,DeMarchi:2011ye}. Further, all the expressions of these kinematic indicators of Taiji triangle for the optimal value (\ref{equ1.1}) are presented.

In the above analysis on the unperturbed Keplerian orbits of SCs, only the contribution from the gravitational field of the Sun is taken into account.  For a more accurate analysis on the relative motion between SCs, which is very important in the implementation of TDI~\cite{Pucacco:2010mn}, the contributions of the Earth,
the Moon, and other planets should be taken into consideration.  In the present paper, we focus on analyzing the perturbation of the Earth on SCs as did for LISA in Ref.~\cite{Dhurandhar:2008yu} and enhance the precision of the orbits to the order of $e^3$.

The ecliptic plane is chosen as the $x$-$y$ plane of the heliocentric coordinate system, and the direction of $x$
axis may be chosen arbitrarily relative to the major axis of the Earth's orbit. By adjusting the mean anomaly of the Earth Kepler's equation, the nominal trailing angle of Taiji constellation following the Earth is set to be $\chi(\approx\pm\pi/9)$ from the viewpoint of the Sun.  (The negative value of $\chi$ means that the constellation is preceding the Earth.)  According to the equations of motion in the problem of three bodies~\cite{moulton1960}, the influence of the Earth perturbation on three SCs can be determined accurately in the heliocentric coordinate system.  Compared with the previous discussion on the Earth perturbation on LISA SCs~\cite{Dhurandhar:2008yu,Pucacco:2010mn}, our results include the influence of the interaction between the Sun and the Earth and the effect of the eccentricity $e'$ of the Earth's orbit. Our calculation shows that
the perturbative solutions of the leading order, the same as $e^2$ order, take the same form for three SCs.  Therefore, they do not affect the relative motion between SCs.  In other words, they do not contribute to the  variations of the kinematic indicators of Taiji triangle.  Even though, they lead to the change of the above trailing angle of Taiji constellation. As to the perturbative solutions of the next leading order, whose orders are the same as $e^{3}$, they have the contributions to the relative motion between SCs.  The analytic expressions for the leading-order corrections to all the kinematic indicators of Taiji triangle resulted from the Earth and the modified expression for the above trailing angle of Taiji constellation to the order of $e^3$ are first presented. In these expressions, the contribution of the tiny difference between the average angular velocities of the Earth and SCs are also considered because the semi-major axes of the orbits of the Earth and SCs are the same but the total mass of the Sun and the Earth is different from that of the Sun and each SC. Since the difference is very tiny, the Earth perturbs SCs in an almost resonance way, so that the terms characterized by $\Omega(t-t_{0})$ and $\Omega^2(t-t_{0})^2$ exist in the results related to the perturbation of the Earth, which will result in unbounded growing of the perturbations over time, as indicated by Ref.~\cite{Dhurandhar:2008yu}. Therefore, if one wishes to lengthen the running time, Taiji configuration needs to be restored to their initial state after about $3$ years.

The paper is arranged as follows.   In the next section, the unperturbed Keplerian orbits of SCs are analyzed.
The perturbation of the Earth on SCs is studied in section 3.  In the last section, we shall make some concluding remarks.   In the present paper, no summation is taken for repeated indices.
\section{Orbit analysis of SCs\label{Sec:second}}
\subsection{Unperturbed Keplerian orbits of SCs\label{Sec:second1}}
There are more than one model to choose for the orbit design of SCs. In view of the similarity between Taiji program and LISA mission, we adopt the model in Refs. \cite{Dhurandhar:2004rv,Nayak:2006zm}, designed for LISA mission originally, as one part of the pre-study of Taiji program.

The heliocentric coordinate system with coordinates $\{x,y,z\}$ is chosen as follows:
\begin{itemize}
\item The origin is located at the center of mass of the Sun;
\item $(x,y,z)$ are the right-handed Cartesian coordinates, where the $x$-$y$ plane is the ecliptic plane.
\end{itemize}
The radius vector of SC{$k$} ($k=1,2,3$) in the heliocentric coordinate system is $\boldsymbol{r}_{k}=(x_k,y_k,z_k)$. $\boldsymbol{r}_{1}$ is chosen by
\begin{equation}\label{equ2.1}
\left\{\begin{array}{l}
\displaystyle x_{1}=R(e+\cos\psi_{1})\cos\varepsilon,\smallskip\\
\displaystyle y_{1}=R\sqrt{1-e^2}\sin\psi_{1},\smallskip\\
\displaystyle z_{1}=R(e+\cos\psi_{1})\sin\varepsilon,
\end{array}\right.
\end{equation}
where $\varepsilon$ is the inclination of the orbit of SC$1$ with respect to the ecliptic plane, and $\psi_{1}$ is the eccentric anomaly of SC$1$. $\psi_{1}$ satisfies Kepler's equation
\begin{equation}\label{equ2.2}
\psi_{1}+e\sin\psi_{1}=\Omega t,
\end{equation}
where $\Omega$ is the average angular velocity of SC1. Moreover, the inclination $\varepsilon$ satisfies \cite{Nayak:2006zm}
\begin{equation}\label{equ2.3}
\left\{\begin{array}{l}
\displaystyle \cos\varepsilon=\frac{\sqrt{3}}3 \frac{\sqrt{3}+2\alpha\cos{\phi}}{1+e},\smallskip\\
\displaystyle \sin\varepsilon=\frac{\sqrt{3}}3\frac{2\alpha \sin{\phi}}{1+e} ,
\end{array}\right.
\end{equation}
where $\alpha$ is the small parameter for the expansion in Refs.~\cite{Dhurandhar:2004rv,Nayak:2006zm}, related to the constant angle $\phi$ by
\begin{equation}\label{equ2.4}
\alpha=\frac{\sqrt{3}}{2}\left(\sqrt{e^2+2e+\cos^2{\phi}}-\cos{\phi}\right)>0.
\end{equation}
The orbits of SC2 and SC3 are obtained, respectively, by rotating that of SC1 by $2\pi/3,4\pi/3$ about the $z$ axis, where their phases also need to be adjusted correspondingly~\cite{Dhurandhar:2004rv,Nayak:2006zm}, i.e.
$\boldsymbol{r}_{2},\boldsymbol{r}_{3}$ satisfy
\begin{figure}
\centering
\includegraphics[scale=0.5]{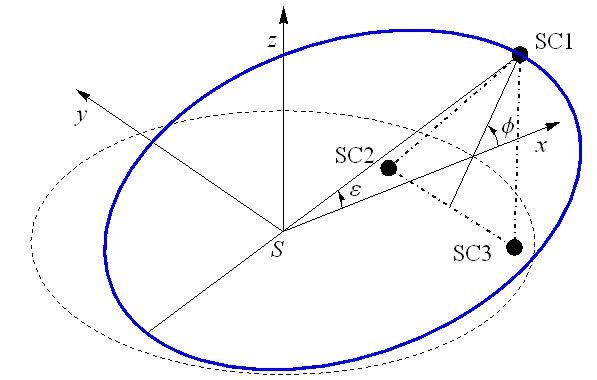}
\caption{\label{fig:1} Plot of the geometry of Taiji configuration. The heliocentric coordinate system is labeled by $(x,y,z)$. SC$k\ (k=1,2,3)$ denote three spacecrafts, respectively, and $S$ denotes the Sun. The dashed circle is in the ecliptic plane with radius $R$.}
\label{fig1}
\end{figure}
\begin{equation}\label{equ2.5}
\left\{\begin{array}{l}
\displaystyle x_{2}=R(e+\cos\psi_{2})\cos\varepsilon\cos{\frac{2\pi}{3}}-R\sqrt{1-e^2}\sin\psi_{2}\sin{\frac{2\pi}{3}},\smallskip\\
\displaystyle y_{2}=R(e+\cos\psi_{2})\cos\varepsilon\sin{\frac{2\pi}{3}}+R\sqrt{1-e^2}\sin\psi_{2}\cos{\frac{2\pi}{3}},\smallskip\\
\displaystyle z_{2}=R(e+\cos\psi_{2})\sin\varepsilon,
\end{array}\right.
\end{equation}
\begin{equation}\label{equ2.6}
\left\{\begin{array}{l}
\displaystyle x_{3}=R(e+\cos\psi_{3})\cos\varepsilon\cos{\frac{4\pi}{3}}-R\sqrt{1-e^2}\sin\psi_{3}\sin{\frac{4\pi}{3}},\smallskip\\
\displaystyle y_{3}=R(e+\cos\psi_{3})\cos\varepsilon\sin{\frac{4\pi}{3}}+R\sqrt{1-e^2}\sin\psi_{3}\cos{\frac{4\pi}{3}},\smallskip\\
\displaystyle z_{3}=R(e+\cos\psi_{3})\sin\varepsilon,
\end{array}\right.
\end{equation}
and the eccentric anomalies $\psi_2,\psi_3$ of SC2, SC3 satisfy \cite{Nayak:2006zm}, respectively,
\begin{eqnarray}
\label{equ2.7}\psi_{2}+e\sin\psi_{2}=\Omega t-\frac{2\pi}{3},\\
\label{equ2.8}\psi_{3}+e\sin\psi_{3}=\Omega t-\frac{4\pi}{3}.
\end{eqnarray}
The geometry of the Taiji configuration is presented in FIG.~\ref{fig1}. Equations~(\ref{equ2.3}) and (\ref{equ2.4}) imply that both $\varepsilon$ and $\phi$ have the same sign. As is mentioned above, $\varepsilon$ is the inclination of the orbits of SCs with respect to the ecliptic plane, and thus, $\pm|\varepsilon|$ can provide two kinds of Taiji configurations which are symmetry about the ecliptic plane.
\subsection{Expansions of the unperturbed Keplerian orbits of SCs to $e^3$ order\label{Sec:second2}}
Since the eccentricity $e \ll 1$, the Kepler's equations (\ref{equ2.2}), (\ref{equ2.7}), and (\ref{equ2.8}), as the transcendental equations, can be dealt with by the iterative method, and then, the combination of Eqs.~(\ref{equ2.1}) and (\ref{equ2.3})---(\ref{equ2.6}) can bring about the expansion of the unperturbed Keplerian orbit of SC$k$, namely $\boldsymbol{r}_{k}$, to $e^3$ order:
\begin{equation}\label{equ2.16}
\boldsymbol{r}_{k}=\boldsymbol{r}_{k}^{(0)}+\boldsymbol{r}_{k}^{(1)}e+\boldsymbol{r}_{k}^{(2)}e^2
+\boldsymbol{r}_{k}^{(3)}e^3, \mbox{\ for\ } k=1,2,3,
\end{equation}
where $\boldsymbol{r}_{k}^{(n)}$$=$$(x_{k}^{(n)},y_{k}^{(n)},z_{k}^{(n)})$ $(n=0,1,2,3)$. The detailed derivation and the expressions of $x_{k}^{(n)},y_{k}^{(n)},$ and $z_{k}^{(n)}$ are all put in Appendix A.
The barycentre of three SCs can be derived by
\begin{equation}\label{equ2.21}
\boldsymbol{r}=\frac{1}{3}(\boldsymbol{r}_{1}+\boldsymbol{r}_{2}+\boldsymbol{r}_{3})=(x, y, z)
\end{equation}
with
\begin{equation}\label{equ2.22}
\left\{\begin{array}{l}
\displaystyle x=R\left[\cos(\Omega t)-\Big(\frac{1}{2}+\frac{1}{4}\tan^2\phi\Big)\cos(\Omega t)e^2+\left(\Big(\frac{1}{2}\tan^2\phi+\frac{1}{4}\tan^4\phi\Big)\cos(\Omega t)\right.\right.\smallskip\\
\displaystyle\phantom{x=}\qquad\left.\left.-\Big(\frac{1}{24}-\frac{1}{8}\tan^2\phi\Big)\cos(2\Omega t)-\frac{1}{3}\cos(4\Omega t)\right)e^3\right],\smallskip\\
\displaystyle y=R\left[\sin(\Omega t)-\Big(\frac{1}{2}+\frac{1}{4}\tan^2\phi\Big)\sin(\Omega t)e^2+\left(\Big(\frac{1}{2}\tan^2\phi+\frac{1}{4}\tan^4\phi\Big)\sin(\Omega t)\right.\right.\smallskip\\
\displaystyle\phantom{x=}\qquad\left.\left.+\Big(\frac{1}{24}-\frac{1}{8}\tan^2\phi\Big)\sin(2\Omega t)-\frac{1}{3}\sin(4\Omega t)\right)e^3\right],\smallskip\\
\displaystyle z=R\tan\phi\left[\frac{3}{2}e^2-\left(\frac{3}{2}+\frac{3}{4}\tan^2\phi-\frac{3}{8}\cos(3\Omega t)\right)e^3\right]\smallskip.
\end{array}\right.
\end{equation}
Clearly, up to $e^1$ order, the trajectory of the barycentre of three SCs is a circle in the ecliptic plane with a radius of $R$, and moreover, the fact of $z\neq0$ means that the barycentre of three SCs is not always in the ecliptic plane.
\subsection{Relative motion between SCs\label{Sec:second3}}
Firstly, set $\boldsymbol{n}:=(\boldsymbol{r}_{1}-\boldsymbol{r}_{2})\times(\boldsymbol{r}_{2}-\boldsymbol{r}_{3})$, and $\boldsymbol{n}$ is the normal vector of Taiji plane. The tilt angle can be calculated by
\begin{equation}\label{equ2.23}
\pm\arccos\left(\left|\frac{\boldsymbol{n}}{|\boldsymbol{n}|}\cdot (0,0,1)\right|\right)=\phi+O(e),
\end{equation}
where the signs ``$\pm$'' on the left hand side of the above equation represent two cases of $\pm|\phi|$, and they are equivalent to the cases of $\pm|\varepsilon|$ mentioned before, respectively. Obviously, up to $e^0$ order, the tilt angle remains the constant angle $\phi$, and that is to say, $\phi$ is the leading-order term of the tilt angle.

Next, the relative radius vectors of SCs, namely $\boldsymbol{r}_{ij}:=\boldsymbol{r}_{i}-\boldsymbol{r}_{j}\ (i,j=1,2,3,i\neq j)$, to $e^3$ order can be derived by Eqs.~(\ref{equ2.16}) and (\ref{equ2.17}):
\begin{equation}\label{equ2.24}
\boldsymbol{r}_{ij}=\boldsymbol{r}_{ij}^{(1)}e+\boldsymbol{r}_{ij}^{(2)}e^2+\boldsymbol{r}_{ij}^{(3)}e^3,
\end{equation}
where $\boldsymbol{r}_{ij}^{(n)}=\boldsymbol{r}_{i}^{(n)}-\boldsymbol{r}_{j}^{(n)}\ (n=1,2,3)$, and their expressions can be obtained by Eqs.~(\ref{equ2.18})---(\ref{equ2.20}). The arm-length of SC$i$ and SC$j$ to $e^3$ order is
\begin{eqnarray}
l_{ij}&:=&\sqrt{\boldsymbol{r}_{ij}\cdot\boldsymbol{r}_{ij}}=R\left(a_{ij}^{(2)}e^2+a_{ij}^{(3)}e^3+a_{ij}^{(4)}e^4+\cdots\right)^{\frac{1}{2}}\nonumber\\
\label{equ2.26}&&=R\left[\left(\sqrt{a_{ij}^{(2)}}\right)e+\frac{a_{ij}^{(3)}}{2\sqrt{a_{ij}^{(2)}}}e^2+\left(\frac{a_{ij}^{(4)}}{2\sqrt{a_{ij}^{(2)}}}-\frac{\left(a_{ij}^{(3)}\right)^2}{8\left(a_{ij}^{(2)}\right)^{3/2}}\right)e^3\right],
\end{eqnarray}
where
\begin{equation}\label{equ2.27}
a_{ij}^{(2)}=\frac{\boldsymbol{r}_{ij}^{(1)}\cdot \boldsymbol{r}_{ij}^{(1)}}{R^2},\quad
a_{ij}^{(3)}=\frac{2\boldsymbol{r}_{ij}^{(1)}\cdot \boldsymbol{r}_{ij}^{(2)}}{R^2},\quad
a_{ij}^{(4)}=\frac{\boldsymbol{r}_{ij}^{(2)}\cdot \boldsymbol{r}_{ij}^{(2)}+2\boldsymbol{r}_{ij}^{(1)}\cdot \boldsymbol{r}_{ij}^{(3)}}{R^2}.
\end{equation}
The expressions for $a_{ij}^{(n)}$ ($n=2, 3, 4$) are, respectively,
\begin{equation}\label{equ2.28}
\left\{\begin{array}{l}
\displaystyle a_{ij}^{(2)}=\frac{15}{2}+\frac{3}{2}\tan^2\phi+\left(\frac{9}{2}-\frac{3}{2}\tan^2\phi\right)\cos(2\theta_{ij}),\smallskip\\
\displaystyle a_{ij}^{(3)}=-3\tan^2\phi-\frac{3}{2}\tan^4\phi-\frac{15}{4}\tan^2\phi\cos\theta_{ij}+\left(3\tan^2\phi+\frac{3}{2}\tan^4\phi\right)\cos(2\theta_{ij})+\left(-3+\frac{3}{4}\tan^2\phi\right)\cos(3\theta_{ij}),\smallskip\\
\displaystyle a_{ij}^{(4)}=-\frac{21}{32}-\frac{57}{16}\tan^2\phi+\frac{75}{16}\tan^4\phi+\frac{15}{8}\tan^6\phi+\left(\frac{15}{2}\tan^2\phi+\frac{15}{4}\tan^4\phi\right)\cos\theta_{ij}
+\left(-\frac{27}{8}-\frac{27}{4}\tan^2\phi\right.\smallskip\\
\displaystyle \phantom{a_{ij4}=}\left.-\frac{9}{2}\tan^4\phi-\frac{15}{8}\tan^6\phi\right)\cos(2\theta_{ij})+\left(-\frac{3}{2}\tan^2\phi-\frac{3}{4}\tan^4\phi\right)\cos(3\theta_{ij})+\left(-\frac{9}{32}+\frac{3}{16}\tan^2\phi\right)\cos(4\theta_{ij})
\end{array}\right.
\end{equation}
with
\begin{equation}\label{equ2.29}
\left\{\begin{array}{l}
\displaystyle \theta_{12}=\Omega t-\frac{\pi}{3},\smallskip\\
\displaystyle \theta_{23}=\Omega t-\pi,\smallskip\\
\displaystyle \theta_{31}=\Omega t-\frac{5\pi}{3}.
\end{array}\right.
\end{equation}
It is easy to know that the three arm-lengths of Taiji triangle change periodically over time, and they depend on the angle $\phi$ and are not equal to each other in general. Further, Eqs.~(\ref{equ2.26})---(\ref{equ2.29}) show that $l_{23}$ and $l_{31}$ are only the phase-shifted versions of $l_{12}$, which is resulted from the symmetry in the previous orbit model of SCs. If $\phi=\pm\pi/3$, the arm-length of SC$i$ and SC$j$ to $e^3$ order and its corresponding rate of change are, respectively,
\begin{eqnarray}
l_{ij}&=&R\left[2\sqrt{3}e+\left(-\frac{15\sqrt{3}}{8}-\frac{15\sqrt{3}}{16}\cos\theta_{ij}+\frac{15\sqrt{3}}{8}\cos(2\theta_{ij})-\frac{\sqrt{3}}{16}\cos(3\theta_{ij})\right)e^2+\left(\frac{5489\sqrt{3}}{1024}+\frac{1095\sqrt{3}}{256}\cos\theta_{ij}\right.\right.\nonumber\\
\label{equ2.30}&&\left.\left.-\frac{16239\sqrt{3}}{2048}\cos(2\theta_{ij})-\frac{285\sqrt{3}}{512}\cos(3\theta_{ij})-\frac{441\sqrt{3}}{1024}\cos(4\theta_{ij})+\frac{15\sqrt{3}}{512}\cos(5\theta_{ij})-\frac{\sqrt{3}}{2048}\cos(6\theta_{ij})\right)e^3\right],\\
v_{ij}&=&R\Omega\left[0e+\left(\frac{15\sqrt{3}}{16}\sin\theta_{ij}-\frac{15\sqrt{3}}{4}\sin(2\theta_{ij})+\frac{3\sqrt{3}}{16}\sin(3\theta_{ij})\right)e^2+\left(-\frac{1095\sqrt{3}}{256}\sin\theta_{ij}+\frac{16239\sqrt{3}}{1024}\sin(2\theta_{ij})\right.\right.\nonumber\\
\label{equ2.31}&&\left.\left.+\frac{855\sqrt{3}}{512}\sin(3\theta_{ij})+\frac{441\sqrt{3}}{256}\sin(4\theta_{ij})-\frac{75\sqrt{3}}{512}\sin(5\theta_{ij})+\frac{3\sqrt{3}}{1024}\sin(6\theta_{ij})\right)e^3\right].
\end{eqnarray}
Obviously, up to $e^1$ order, the three arm-lengths of Taiji triangle remain $2\sqrt{3}Re$, and their corresponding rates of change remain zero. Therefore, under the cases of $\phi=\pm\pi/3$, Taiji triangle is approximately equilateral one. The vertex angles of Taiji triangle between the relative radius vectors of SCs $\boldsymbol{r}_{ki},\boldsymbol{r}_{kj}\ (i\neq j)$, denoted by $\beta_{ij}$, is defined by
\begin{equation}\label{equ2.32}
\beta_{ij}=\arccos\left(\frac{\boldsymbol{r}_{ki}\cdot \boldsymbol{r}_{kj}}{|\boldsymbol{r}_{ki}| |\boldsymbol{r}_{kj}|}\right).
\end{equation}
A direct calculation shows that up to $e^2$ order,
\begin{eqnarray}
\label{equ2.33}
\beta_{ij}&=&\frac{\pi}{3}+\left(-\frac{15\sqrt{3}}{32}\cos\theta_{ij}+\frac{15\sqrt{3}}{16}\cos(2\theta_{ij})\right)e
+\left(\frac{135\sqrt{3}}{128}\cos\theta_{ij}-\frac{5997\sqrt{3}}{2048}\cos(2\theta_{ij})+\frac{447\sqrt{3}}{1024}\cos(4\theta_{ij})
+\frac{15\sqrt{3}}{512}\cos(5\theta_{ij})\right)e^2.\nonumber\\
\phantom{111}
\end{eqnarray}
It shows that the three vertex angles of Taiji triangle remain $\pi/3$ up to $e^0$ order. The result is compatible with the result of arm-length to $e^1$ order via the cosine theorem in Euclidean geometry. Moreover,
three vertex angles of Taiji triangle are identical to each other up to a phase shift of $2\pi/3$, which further confirms that three SCs are symmetrical in the present orbit model.

\subsection{Optimization of the unperturbed Keplerian orbits of SCs\label{Sec:second4}}

As LISA~\cite{Dhurandhar:2008yu}, Taiji needs to suppress the laser frequency noise below the other secondary noises
by TDI.
The first generation TDI only works well for the stationary Taiji configuration.
As shown in the previous subsection, the higher-order terms of kinematic indicators change periodically over time and thus the Taiji triangle is only approximately equilateral one even under the cases of $\phi=\pm\pi/3$.  The extra moving part in the relative motion between SCs may require modified first generation TDI or further, the second generation TDI~\cite{Dhurandhar:2008yu,Tinto:2003vj,Vallisneri:2005ji,Tinto:2014lxa}. The orbital eccentricity of Taiji SCs is smaller than that of LISA SCs (presented in Refs.~\cite{Dhurandhar:2004rv,Nayak:2006zm,Dhurandhar:2008yu}), and therefore, a reduction of the amplitude of the relative motion between SCs could contribute to considering a simpler TDI strategy for Taiji program, as that for LISA~\cite{Nayak:2006zm,Dhurandhar:2008yu}. Further, the extra relatively moving part will also cause the Doppler shift of the laser frequency, which will interfere with the measurement of GWs~\cite{Nayak:2006zm}, and in order to lower this adverse effect, the amplitude of the extra relatively moving part needs to be minimized, while the relatively stationary part remains the same.
This is the optimization of the orbits of SCs.

If the angle $\phi$ is set to be
\begin{equation}\label{equ2.34}
\phi=\pm\left(\frac{\pi}{3}+\delta\right),
\end{equation}
where the parameter $\delta$ is the same order as $e$, and by Eqs.~(\ref{equ2.30}), (\ref{equ2.31}), and (\ref{equ2.33}), the expansions of $l_{ij}$ and $v_{ij}$ to $e^{m}\delta^{n}\ (m+n=3)$ order and $\beta_{ij}$ to $e^{m}\delta^{n}\ (m+n=2)$ are, respectively,
\begin{eqnarray}
l_{ij}&=&R\left[2\sqrt{3}e+\left(3-3\cos(2\theta_{ij})\right)e\delta
+\left(-\frac{15\sqrt{3}}{8}-\frac{15\sqrt{3}}{16}\cos\theta_{ij}+\frac{15\sqrt{3}}{8}\cos(2\theta_{ij})-\frac{\sqrt{3}}{16}\cos(3\theta_{ij})\right)e^2
+\left(\frac{31\sqrt{3}}{8}\right.\right.\nonumber\\
&&\left.-\frac{7\sqrt{3}}{2}\cos(2\theta_{ij})-\frac{3\sqrt{3}}{8}\cos(4\theta_{ij})\right)e\delta^2
+\left(-\frac{633}{32}-\frac{219}{32}\cos\theta_{ij}+\frac{147}{8}\cos(2\theta_{ij})+\frac{57}{64}\cos(3\theta_{ij})+\frac{45}{32}\cos(4\theta_{ij})\right.\nonumber\\
&&\left.-\frac{3}{64}\cos(5\theta_{ij})\right)e^2\delta
+\left(\frac{5489\sqrt{3}}{1024}+\frac{1095\sqrt{3}}{256}\cos\theta_{ij}-\frac{16239\sqrt{3}}{2048}\cos(2\theta_{ij})-\frac{285\sqrt{3}}{512}\cos(3\theta_{ij})-\frac{441\sqrt{3}}{1024}\cos(4\theta_{ij})\right.\nonumber\\
\label{equ2.35}&&\left.\left.+\frac{15\sqrt{3}}{512}\cos(5\theta_{ij})-\frac{\sqrt{3}}{2048}\cos(6\theta_{ij})\right)e^3\right],\\
v_{ij}&=&R\Omega\left[0e+6\sin(2\theta_{ij})e\delta+\left(\frac{15\sqrt{3}}{16}\sin\theta_{ij}-\frac{15\sqrt{3}}{4}\sin(2\theta_{ij})+\frac{3\sqrt{3}}{16}\sin(3\theta_{ij})\right)e^2+\bigg(7\sqrt{3}\sin(2\theta_{ij})\right.\nonumber\\
&&\left.+\frac{3\sqrt{3}}{2}\sin(4\theta_{ij})\right)e\delta^2+\left(\frac{219}{32}\sin\theta_{ij}-\frac{147}{4}\sin(2\theta_{ij})-\frac{171}{64}\sin(3\theta_{ij})-\frac{45}{8}\sin(4\theta_{ij})+\frac{15}{64}\sin(5\theta_{ij})\right)e^2\delta\nonumber\\
\label{equ2.36}&&+\left(-\frac{1095\sqrt{3}}{256}\sin\theta_{ij}
+\frac{16239\sqrt{3}}{1024}\sin(2\theta_{ij})+\frac{855\sqrt{3}}{512}\sin(3\theta_{ij})+\frac{441\sqrt{3}}{256}\sin(4\theta_{ij})-\frac{75\sqrt{3}}{512}\sin(5\theta_{ij})\right.\nonumber\\
&&\left.\left.+\frac{3\sqrt{3}}{1024}\sin(6\theta_{ij})\right)e^3\right],\\
\beta_{ij}&=&\frac{\pi}{3}+\left(-\frac{3}{2}\cos(2\theta_{ij})\right)\delta+\left(-\frac{15\sqrt{3}}{32}\cos\theta_{ij}+\frac{15\sqrt{3}}{16}\cos(2\theta_{ij})\right)e+\left(-\sqrt{3}\cos(2\theta_{ij})+\frac{3\sqrt{3}}{8}\cos(4\theta_{ij})\right)\delta^2\nonumber\\
&&+\left(-\frac{27}{16}\cos\theta_{ij}+\frac{51}{8}\cos(2\theta_{ij})-\frac{45}{32}\cos(4\theta_{ij})-\frac{3}{64}\cos(5\theta_{ij})\right)\delta e+\left(\frac{135\sqrt{3}}{128}\cos\theta_{ij}-\frac{5997\sqrt{3}}{2048}\cos(2\theta_{ij})\right.\nonumber\\
\label{equ2.37}&&\left.+\frac{447\sqrt{3}}{1024}\cos(4\theta_{ij})+\frac{15\sqrt{3}}{512}\cos(5\theta_{ij})\right)e^2.
\end{eqnarray}
The leading-order terms of the above results being independent on $\delta$ imply that the relatively stationary part remains unchanged. The higher-order terms of the above results  depend on $\delta$, and then,
the amplitude of the extra relatively moving part could be minimized by adjusting the value of $\delta$. Now, only $l_{ij}, v_{ij}$ and $\beta_{ij}$ up to their next-leading-order terms (i.e. the lowest order terms containing the parameter $\delta$) are considered.  They are
\begin{eqnarray}
\label{equ2.38}l_{ij}&=&R\left[2\sqrt{3}e+\left(3-3\cos(2\theta_{ij})\right)e\delta
+\left(-\frac{15\sqrt{3}}{8}-\frac{15\sqrt{3}}{16}\cos\theta_{ij}+\frac{15\sqrt{3}}{8}\cos(2\theta_{ij})-\frac{\sqrt{3}}{16}\cos(3\theta_{ij})\right)e^2\right],\\
\label{equ2.39}v_{ij}&=&R\Omega\left[0e+6\sin(2\theta_{ij})e\delta+\left(\frac{15\sqrt{3}}{16}\sin\theta_{ij}-\frac{15\sqrt{3}}{4}\sin(2\theta_{ij})+\frac{3\sqrt{3}}{16}\sin(3\theta_{ij})\right)e^2\right],\\
\label{equ2.40}\beta_{ij}&=&\frac{\pi}{3}+\left(-\frac{3}{2}\cos(2\theta_{ij})\right)\delta+\left(-\frac{15\sqrt{3}}{32}\cos\theta_{ij}+\frac{15\sqrt{3}}{16}\cos(2\theta_{ij})\right)e.
\end{eqnarray}
Equations.~(\ref{equ2.35})---(\ref{equ2.37}) are optimized when the standard deviation of Eqs.~(\ref{equ2.38})---(\ref{equ2.40}) reach their minimums.

For the quantity $A(t)$, varying periodically over time, its average $\big<A(t)\big>$ within $n\ (n=1,2,3\cdots)$ years is defined as
\begin{eqnarray}
\label{equ2.41}\big<A(t)\big>&:=&\frac{\Omega}{2n\pi}\int_{t_{0}}^{t_{0}+2n\pi/\Omega}A(t)dt.
\end{eqnarray}
Then, the amplitude of $A(t)$, characterized by its standard deviation within $n$ years, is
\begin{eqnarray}
\label{equ2.42}\sigma(A(t)):=\sqrt{\big<\left(A(t)-\big<A(t)\big>\right)^2\big>}.
\end{eqnarray}
The averages of Eqs. (\ref{equ2.38})---(\ref{equ2.40}) within $n$ years are, respectively,
\begin{eqnarray}
\label{equ2.43}\big<l_{ij}\big>=R\left(2\sqrt{3}e+3e\delta-\frac{15\sqrt{3}}{8}e^2\right),\qquad \big<v_{ij}\big>=0,\qquad\big<\beta_{ij}\big>=\frac{\pi}{3},
\end{eqnarray}
and further by Eq.~(\ref{equ2.42}), their standard deviations within $n$ years are, respectively,
\begin{eqnarray}
\label{equ2.44}\sigma(l_{ij})&=&\frac{\sqrt{6}Re^2}{32}\left[226+\left(16\sqrt{3}\right)^2\left(\tilde{\delta}-\frac{5\sqrt{3}}{8}\right)^2\right]^{\frac{1}{2}},\\
\label{equ2.45}\sigma(v_{ij})&=&\frac{3\sqrt{6}R\Omega e^2}{32}\left[26+\left(\frac{32\sqrt{3}}{3}\right)^2\left(\tilde{\delta}-\frac{5\sqrt{3}}{8}\right)^2\right]^{\frac{1}{2}},\\
\label{equ2.46}\sigma(\beta_{ij})&=&\frac{3\sqrt{2}e}{4}\left[\frac{75}{256}+\left(\tilde{\delta}-\frac{5\sqrt{3}}{8}\right)^2\right]^{\frac{1}{2}},
\end{eqnarray}
where $\tilde{\delta}:=\delta/e$. Obviously, when $\tilde{\delta}=5\sqrt{3}/8$, $\sigma(l_{ij}), \sigma(v_{ij})$ and $\sigma(\beta_{ij})$  take the minimums.  Namely, when
\begin{equation}\label{equ2.47}
\delta=\frac{5\sqrt{3}e}{8}\Leftrightarrow\phi=\pm\left(\frac{\pi}{3}+\frac{5\sqrt{3}e}{8}\right),
\end{equation}
the amplitude of the extra relatively moving part between SCs is minimized. Thus, the change of the kinematic indicators of Taiji triangle can all be suppressed effectively. Compared with the original case, namely the case of $\tilde{\delta}=0$, the ratios of reduction are, respectively,
\begin{eqnarray}
\label{equ2.48}\frac{\sigma(l_{ij})|_{\tilde{\delta}=0}}{\sigma(l_{ij})|_{\tilde{\delta}=5\sqrt{3}/8}}\approx2.23,\qquad
\frac{\sigma(v_{ij})|_{\tilde{\delta}=0}}{\sigma(v_{ij})|_{\tilde{\delta}=5\sqrt{3}/8}}\approx4.05,\qquad
\frac{\sigma(\beta_{ij})|_{\tilde{\delta}=0}}{\sigma(\beta_{ij})|_{\tilde{\delta}=5\sqrt{3}/8}}\approx2.24.
\end{eqnarray}
Finally, applying the result~(\ref{equ2.47}) into the Eqs.~(\ref{equ2.35})---(\ref{equ2.37}), the corresponding expressions of $l_{ij}$ and $v_{ij}$ to $e^3$ order and $\beta_{ij}$ to $e^2$ order are obtained, respectively,
\begin{eqnarray}
\label{equ2.49}l_{ij}&=&R\left[2\sqrt{3}e+\left(-\frac{15\sqrt{3}}{16}\cos\theta_{ij}-\frac{\sqrt{3}}{16}\cos(3\theta_{ij})\right)e^2
+\left(-\frac{2521\sqrt{3}}{1024}-\frac{1119\sqrt{3}}{2048}\cos(2\theta_{ij})+\frac{9\sqrt{3}}{1024}\cos(4\theta_{ij})-\frac{\sqrt{3}}{2048}\cos(6\theta_{ij})\right)e^3\right]\nonumber\\
&&\phantom{111}\\
\label{equ2.50}v_{ij}&=&R\Omega\left[\left(\frac{15\sqrt{3}}{16}\sin\theta_{ij}+\frac{3\sqrt{3}}{16}\sin(3\theta_{ij})\right)e^2
+\left(\frac{1119\sqrt{3}}{1024}\sin(2\theta_{ij})-\frac{9\sqrt{3}}{256}\sin(4\theta_{ij})+\frac{3\sqrt{3}}{1024}\sin(6\theta_{ij})\right)e^3\right],\\
\label{equ2.51}\beta_{ij}&=&\frac{\pi}{3}+\left(-\frac{15\sqrt{3}}{32}\cos\theta_{ij}\right)e
+\left(-\frac{237\sqrt{3}}{2048}\cos(2\theta_{ij})-\frac{3\sqrt{3}}{1024}\cos(4\theta_{ij})\right)e^2.
\end{eqnarray}
In Ref.~\cite{Nayak:2006zm}, with the help of Hill system or CW system~\cite{Pucacco:2010mn},
the arm-length $l_{12}$ of LISA triangle and its rate of change $v_{12}$ are expanded to $\alpha^2$ order under the case of $\phi>0$, where $\alpha$ is defined by
Eq.~(\ref{equ2.4}), and when
\begin{equation}\label{equ2.52}
\delta=\frac{5\alpha}{8}\Leftrightarrow\phi=\frac{\pi}{3}+\frac{5\alpha}{8},
\end{equation}
$\sigma(l_{12})$ and $\sigma(v_{12})$ reach their minimums, respectively. Eq.~(\ref{equ2.14}) shows that at $e^1$ order, $\alpha=\sqrt{3}e$, so the result (\ref{equ2.47}) in present paper is compatible with that of LISA in Refs.~\cite{Nayak:2006zm,DeMarchi:2011ye}, namely Eq.~(\ref{equ2.52}).
\section{Influence of the Earth perturbation on SCs\label{Sec:third}}
Besides the non-stationary configuration resulted from the gravitational field of the Sun, Taiji constellation is also affected by the gravitational fields of the Earth and the other celestial bodies. In order to acquire the more accurate knowledge about the relative motion between SCs, the perturbation of these gravitational fields  need to be taken into account. Here, for simplicity, only the perturbation contributed by the Earth on SCs is discussed. Compared with that in Refs. \cite{Dhurandhar:2008yu,Pucacco:2010mn}, the precision of the orbits are enhanced to the order of $e^3$.
\subsection{Earth's orbit}
The radius vector of the Earth in the heliocentric coordinate system is $\boldsymbol{r}'=(x',y',z')$ with
\begin{equation}\label{equ3.1}
\left\{\begin{array}{l}
\displaystyle x'=R(e'+\cos\psi')\cos\varphi-R\sqrt{1-e'^2}\sin\psi'\sin\varphi,\smallskip\\
\displaystyle y'=R(e'+\cos\psi')\sin\varphi+R\sqrt{1-e'^2}\sin\psi'\cos\varphi,\smallskip\\
\displaystyle z'=0,
\end{array}\right.
\end{equation}
where $e'\approx1.672\times10^{-2},\psi'$ are, respectively, the eccentricity and the eccentric anomaly of the Earth's orbit. $\psi'$ satisfies the following Kepler's equation
\begin{equation}\label{equ3.2}
\psi'+e'\sin\psi'=\Omega' t+\chi-\varphi,
\end{equation}
and according to the Kepler's third law, the average angular velocity $\Omega'$ of the Earth can be derived, namely,
\begin{equation}\label{equ3.3}
\Omega'=\sqrt{\frac{G(m_{S}+m_{E})}{R^3}},
\end{equation}
where $m_{S}$ and $m_{E}$ are the masses of the Sun and the Earth, respectively. Similarly, if the masses of SCs
are denoted by $m$, their average angular velocity $\Omega$ can also be obtained,
\begin{equation}\label{equ3.4}
\Omega=\sqrt{\frac{G(m_{S}+m)}{R^3}}.
\end{equation}
Because of $m\lll m_{S}$, the ratio of these two average angular velocities is
\begin{equation}\label{equ3.5}
\frac{\Omega'}{\Omega}=\sqrt{\frac{1+m_{E}/m_{S}}{1+m/m_{S}}}\approx1+\zeta
\end{equation}
with
\begin{equation}\label{equ3.6}
\zeta:=\frac{m_{E}}{2m_{S}}\approx1.020\times10^{-6}\lesssim e'^3.
\end{equation}
When $\psi'=0$, the Earth is at the aphelion,
\begin{eqnarray*}
\boldsymbol{r}'=\left(R(e'+1)\cos\varphi,R(e'+1)\sin\varphi,0\right).
\end{eqnarray*}
It shows that $\varphi$ is nothing but the angle between the major axis of the Earth's orbit and $x$ axis of the heliocentric coordinate system. The angle $\varphi$ can take any value in the interval $[0,2\pi)$. According to the result in Appendix B, the expansion of $\boldsymbol{r}'$ to $e'^3$ order and $\zeta^1$ order can be written as
\begin{equation}\label{equ3.8}
\boldsymbol{r}'=\boldsymbol{r}'^{[0]}+\boldsymbol{r}'^{[1]}+\boldsymbol{r}'^{[2]}+\boldsymbol{r}'^{[3]},
\end{equation}
where $\boldsymbol{r}'^{[n]}=\big(x'^{[n]},y'^{[n]},z'^{[n]}\big)$ $(n=0,1,2,3)$, the superscript numbers in square brackets indicate the order of $e$ or $e'$.  In particular, $\boldsymbol{r}'^{[3]}$ contains the linear terms of $\zeta$ in addition to $e^3$ or $e'^3$ terms.  The same rule also applies to $\boldsymbol{r}$ and $\stackrel{0}{\boldsymbol{r}}_{k}$, $\stackrel{1}{\boldsymbol{r}}_{k}$, etc in next subsections.

In Refs.~\cite{Dhurandhar:2008yu,Pucacco:2010mn}, the Earth's orbit is simplified to be a circle in the ecliptic plane, namely, given by Eq.~(\ref{equ3.9}). From Eqs.~(\ref{equ2.21}) and (\ref{equ2.22}), at $e^0$ order, the trajectory of the barycentre of three SCs is a circle in $x$-$y$ plane with the same radius as
that of $\boldsymbol{r}'^{[0]}$:
\begin{equation}\label{equ3.13}
\boldsymbol{r}^{[0]}=\boldsymbol{r}^{(0)}=(R\cos(\Omega t), R\sin(\Omega t), 0),
\end{equation}
and thus, the trailing angle $\boldsymbol{r}^{[0]}$ following $\boldsymbol{r}'^{[0]}$ is $\chi$, which is the so-called nominal trailing angle of Taiji constellation following the Earth from the viewpoint of the Sun. In this paper, $\chi$ is assumed to be about $\pm\pi/9$, and the negative value of $\chi$ means that the constellation is preceding the Earth.

\subsection{Perturbation of the Earth on SCs}
The Earth perturbation on every SC can be discussed, separately, so it is actually the problem of three bodies and is inherently non-linear. Therefore, the equations of motion in the problem of three bodies
can be used to deal with the effect of the Earth perturbation. The equation of SC$k$ perturbed by the Earth is
\begin{equation}\label{equ3.14}
\frac{d^2\boldsymbol{r}_{k}}{dt^2}+\frac{\mu\boldsymbol{r}_{k}}{r_{k}^3}=\nabla_{k}R_{k},
\end{equation}
where $\mu=G(m_{S}+m)\approx Gm_{S}$, $r_{k}:=|\boldsymbol{r}_{k}|$,
$$\nabla_{k}:=\left(\frac{\partial}{\partial x_{k}},\frac{\partial}{\partial y_{k}},\frac{\partial}{\partial z_{k}}\right)$$
is the gradient operator, and
\begin{equation}\label{equ3.15}
R_{k}:=Gm_{E}\left(\frac{1}{|\boldsymbol{r}'-\boldsymbol{r}_{k}|}-\frac{\boldsymbol{r}'\cdot\boldsymbol{r}_{k}}{r'^3}\right)
\end{equation}
is the corresponding perturbative function~\cite{moulton1960} with $r':=|\boldsymbol{r}'|$. Thus, Eq.~(\ref{equ3.14}) can be rewritten into
\begin{equation}\label{equ3.16}
\frac{d^2\boldsymbol{r}_{k}}{dt^2}+\frac{\mu\boldsymbol{r}_{k}}{r_{k}^3}=Gm_{E}\left(\frac{\boldsymbol{r}'-\boldsymbol{r}_{k}}{|\boldsymbol{r}'-\boldsymbol{r}_{k}|^3}-\frac{\boldsymbol{r}'}{r'^3}\right).
\end{equation}
Clearly, the first term on the right hand side of above equation is the gravitational acceleration of SC$k$ due to the Earth, and the second term is the negative of the gravitational acceleration of the Sun due to the Earth, which represents the contribution of the interaction between the Sun and the Earth.

Because $m_{E}\ll m_{S}$, the solution of Eq.~(\ref{equ3.16}) can be assumed to have the form of
\begin{equation}\label{equ3.17}
\boldsymbol{r}_{k}=\stackrel{0}{\boldsymbol{r}}_{k}+\stackrel{1}{\boldsymbol{r}}_{k}.
\end{equation}
Here, $\stackrel{0}{\boldsymbol{r}}_{k}$ is the unperturbed Keplerian orbit of SC$k$ in the previous section, and $\stackrel{1}{\boldsymbol{r}}_{k}$ is the perturbative solution.
The equation of $\stackrel{1}{\boldsymbol{r}}_{k}$ is derived in Appendix C and reads
\begin{widetext}
\begin{equation}\label{equ3.20}
\frac{d^2\stackrel{1}{\boldsymbol{r}}_{k}}{dt^2}+\frac{\mu\stackrel{1}{\boldsymbol{r}}_{k}}{\big(\stackrel{0}{r}_{k}\big)^3}
-\frac{3\mu(\stackrel{0}{\boldsymbol{r}}_{k}\cdot\stackrel{1}{\boldsymbol{r}}_{k})\stackrel{0}{\boldsymbol{r}}_{k}}
{\big(\stackrel{0}{r}_{k}\big)^5}=
Gm_{E}\left(\frac{\boldsymbol{r}'-\stackrel{0}{\boldsymbol{r}}_{k}}{\big|\boldsymbol{r}'-\stackrel{0}{\boldsymbol{r}}_{k}\big|^3}
-\frac{\boldsymbol{r}'}{r'^3}\right),
\end{equation}
\end{widetext}
where $\stackrel{0}{r}_{k}=|\stackrel{0}{\boldsymbol{r}}_{k}|$.  The initial condition for the perturbative solution is
\begin{equation}\label{equ3.27}
\stackrel{1}{\boldsymbol{r}}_{k}|_{t=t_0}=0,\qquad \frac{d\stackrel{1}{\boldsymbol{r}}_{k}}{dt}\Big|_{t=t_0}=0,
\end{equation}
where $t_{0}$ is the time of SC$k$ entering the unperturbed Keplerian orbit.  Further, the perturbative solution $\stackrel{1}{\boldsymbol{r}}_{k}$ is assumed to be expanded in terms of the order of $e$,
\begin{equation}\label{equ3.24}
\stackrel{1}{\boldsymbol{r}}_{k}=\stackrel{1}{\boldsymbol{r}}_{k}^{[2]}
+\stackrel{1}{\boldsymbol{r}}_{k}^{[3]}+\cdots,
\end{equation}
where $\stackrel{1}{\boldsymbol{r}}_{k}^{[2]}$ and $\stackrel{1}{\boldsymbol{r}}_{k}^{[3]}$ are governed by Eqs.~(\ref{equ3.25}) and (\ref{equ3.30}), and can be solved with the help of the initial conditions of Eq.~(\ref{equ3.20}), namely, Eq.~(\ref{equ3.27}).

The solution of $\stackrel{1}{\boldsymbol{r}}_{k}^{[2]}
=\Big(\stackrel{1}{x}_{k}^{[2]},\stackrel{1}{y}_{k}^{[2]},\stackrel{1}{z}_{k}^{[2]}\Big)$ is presented by Eq.~(\ref{equ3.29}), and it shows that the magnitude of $\stackrel{1}{\boldsymbol{r}}_{k}^{[2]}$ is determined by the perturbative parameter $\kappa/\lambda^{\frac{3}{2}}\approx7.258\times10^{-5}$ with $\kappa:=m_{E}/m_{S}=2\zeta$ and
\begin{equation}\label{equ3.26}
\lambda:=\left|\frac{\boldsymbol{r}'^{[0]}-\stackrel{0}{\boldsymbol{r}}_{k}^{[0]}}{R}\right|^2
=4\sin^2\left(\frac{\chi}{2}\right)\approx0.1206,
\end{equation}
where $\stackrel{0}{\boldsymbol{r}}_{k}^{[0]}$ is the leading-order term of $\stackrel{0}{\boldsymbol{r}}_{k}$. Obviously, $\kappa/\lambda^{\frac{3}{2}}$ is the same order as $e^2\approx3.351\times10^{-5}$.
It can be proved that $\kappa/\lambda^{\frac{3}{2}}$ is the same as the definitions of the perturbative parameters in Refs.~\cite{Pucacco:2010mn} and \cite{Dhurandhar:2008yu}.
Moreover, $\stackrel{1}{\boldsymbol{r}}_{k}^{[2]}$ have the same expression for $k=1,2,3$.  In other words,  the perturbative solutions of the leading order are the same for three SCs. Therefore, the relative motion between SCs is not affected by $\stackrel{1}{\boldsymbol{r}}_{k}^{[2]}$, and all the kinematic indicators of Taiji triangle are not affected either. But, this solution can change the barycentre of three SCs at $\kappa/\lambda^{\frac{3}{2}}\sim e^2$ order, so the trailing angle of Taiji configuration will be corrected by $\stackrel{1}{\boldsymbol{r}}_{k}^{[2]}$.

As the next-leading-order perturbative solution, $\stackrel{1}{\boldsymbol{r}}_{k}^{[3]}=({\stackrel{1}{x}_{k}^{[3]},} {\stackrel{1}{y}_{k}^{[3]},} {\stackrel{1}{z}_{k}^{[3]}})$ is presented by
\begin{equation}\label{equ3.33}
\left\{\begin{array}{l}
\displaystyle \stackrel{1}{x}_{k}^{[3]}=\frac{\kappa }{\lambda^{\frac{3}{2}}}R\left(E_{xk}e+E_{xk}'e'+\Lambda_{xk}\lambda^{\frac{3}{2}}\right),\smallskip\\
\displaystyle \stackrel{1}{y}_{k}^{[3]}=\frac{\kappa }{\lambda^{\frac{3}{2}}}R\left(E_{yk}e+E_{yk}'e'+\Lambda_{yk}\lambda^{\frac{3}{2}}\right),\smallskip\\
\displaystyle \stackrel{1}{z}_{k}^{[3]}=\frac{\kappa }{\lambda^{\frac{3}{2}}}R\left(E_{zk}e+E_{zk}'e'+\Lambda_{zk}\lambda^{\frac{3}{2}}\right),
\end{array}\right.
\end{equation}
where the expressions of $E_{xk}$, $E_{yk}$, $E_{zk}$, $E_{xk}'$, $E_{yk}'$, $E_{zk}'$, $\Lambda_{xk}$, $\Lambda_{yk}$, and $\Lambda_{zk}$ are shown by Eqs.~(\ref{equ3.34})---(\ref{equ3.42}), respectively. The equation $\lambda^{\frac{3}{2}}\approx4.189\times10^{-2}$ shows $\lambda^{\frac{3}{2}}\gtrsim e'\gtrsim e$, so the order of $\stackrel{1}{\boldsymbol{r}}_{k}^{[3]}$ is the same as $e^3\approx1.940\times10^{-7}$.
Clearly, the above solutions contain three parts, which are characterized by $e,e'$ and $\lambda^{\frac{3}{2}}$, respectively. Among these three parts, only the one related to $e$ depends on $k$.  It is this part that leads to the perturbative solutions of the next leading order for three SCs being different from each other.  That is to say, the part of the next-leading-order perturbative solutions related to $e$ affects the relative motion between SCs and further affects all the kinematic indicators of Taiji triangle.  Moreover, it can be shown that the part of Eq.~(\ref{equ3.33}) related to $e$ does not affect the barycentre of three SCs.  In contrast, the other two parts of the next-leading-order perturbative solutions have no contribution to the relative motion between SCs but can change the barycentre of three SCs at $\kappa e/\lambda^{\frac{3}{2}}\sim \kappa e'/\lambda^{\frac{3}{2}}\sim \kappa\sim e^3$ order in comparison with the leading-order perturbative solution. Therefore, the trailing angle of Taiji configuration will be corrected only by these two parts of the next-leading-order perturbative solutions.
\subsection{Influence of the Earth perturbation on Taiji configuration}
Eqs.~(\ref{equ3.17}) and (\ref{equ3.24}) provide the radius vector of SC$k$ perturbed by the Earth to the order of $e^3$ in heliocentric coordinate system,
\begin{equation}\label{equ3.44}
\boldsymbol{r}_{k}=\stackrel{0}{\boldsymbol{r}}_{k}+\stackrel{1}{\boldsymbol{r}}_{k}^{[2]}
+\stackrel{1}{\boldsymbol{r}}_{k}^{[3]},
\end{equation}
where $\stackrel{0}{\boldsymbol{r}}_{k}$
is given by Eq.~(\ref{equ2.16}). Then, the corresponding relative radius vectors of SCs, namely $\boldsymbol{r}_{ij}:=\boldsymbol{r}_{i}-\boldsymbol{r}_{j}\ (i,j=1,2,3,i\neq j)$,  can be derived:
\begin{equation}\label{equ3.46}
\boldsymbol{r}_{ij}=\stackrel{0}{\boldsymbol{r}}_{ij}+\stackrel{1}{\boldsymbol{r}}_{ij}^{[2]}+\stackrel{1}{\boldsymbol{r}}_{ij}^{[3]},
\end{equation}
where $\stackrel{0}{\boldsymbol{r}}_{ij}$
has been presented in Eq~(\ref{equ2.24}). By Eqs.~(\ref{equ3.29}), (\ref{equ3.33}), and (\ref{equ3.34})---(\ref{equ3.43}),
\begin{eqnarray}
\label{equ3.48}\stackrel{1}{\boldsymbol{r}}_{ij}^{[2]}&=&\stackrel{1}{\boldsymbol{r}}_{i}^{[2]}-\stackrel{1}{\boldsymbol{r}}_{j}^{[2]}=0,\\
\label{equ3.49}\stackrel{1}{\boldsymbol{r}}_{ij}^{[3]}&=&\stackrel{1}{\boldsymbol{r}}_{i}^{[3]}-\stackrel{1}{\boldsymbol{r}}_{j}^{[3]}=\frac{\kappa e }{\lambda^{\frac{3}{2}}}R\boldsymbol{E}_{ij}
\end{eqnarray}
with $\boldsymbol{E}_{ij}:=(E_{xi}-E_{xj},E_{yi}-E_{yj},E_{zi}-E_{zj})$. By use of these results, the arm-length of SC$i$ and SC$j$ perturbed by the Earth and its rate of change to the order of $e^3$ are derived, respectively,
\begin{eqnarray}
\label{equ3.50}l_{ij}&=&\stackrel{0}{l}_{ij}+\stackrel{1}{l}_{ij}^{[3]}=\stackrel{0}{l}_{ij}+\frac{\kappa e }{\lambda^{\frac{3}{2}}}R\frac{\stackrel{0}{\boldsymbol{r}}_{ij}^{[1]}\cdot\boldsymbol{E}_{ij}}{|\stackrel{0}{\boldsymbol{r}}_{ij}^{[1]}|},\\
\label{equ3.51}v_{ij}&=&\stackrel{0}{v}_{ij}+\stackrel{1}{v}_{ij}^{[3]}=\stackrel{0}{v}_{ij}+\frac{\kappa e }{\lambda^{\frac{3}{2}}}R \frac{d}{dt}\left(\frac{\stackrel{0}{\boldsymbol{r}}_{ij}^{[1]}\cdot\boldsymbol{E}_{ij}}{|\stackrel{0}{\boldsymbol{r}}_{ij}^{[1]}|}\right).
\end{eqnarray}
$\stackrel{0}{l}_{ij}$ is the arm-length of SC$i$ and SC$j$ only attracted by the Sun, and $\stackrel{0}{v}_{ij}$ is its corresponding rate of change. Under the optimized case~(\ref{equ2.47}) of
unperturbed Keplerian orbits of SCs,
the expressions of $\stackrel{0}{l}_{ij}$ and $\stackrel{0}{v}_{ij}$ are Eqs.~(\ref{equ2.49}) and (\ref{equ2.50}), respectively. $\stackrel{1}{l}_{ij}^{[3]}$ and $\stackrel{1}{v}_{ij}^{[3]}$ are the leading-order corrections to
$\stackrel{0}{l}_{ij}$ and $\stackrel{0}{v}_{ij}$ due to the attraction of the Earth, respectively, and their expressions can be derived by Eqs.~(\ref{equ3.46})---(\ref{equ3.49}),
\begin{eqnarray}
\stackrel{1}{l}_{ij}^{[3]}&=&\frac{\kappa e }{\lambda^{\frac{3}{2}}}R\left[\frac{125\sqrt{3}}{32}+\frac{147\sqrt{3}}{32}\cos\chi+\frac{123\sqrt{3}}{32}\cos(2\theta_{ij})-\frac{13\sqrt{3}}{4}\cos(\theta_{ij}-\theta_{ij0})
-\frac{21\sqrt{3}}{32}\cos(2\theta_{ij}-2\theta_{ij0})\right.\nonumber\\
&&-\frac{\sqrt{3}}{8}\cos(3\theta_{ij}-\theta_{ij0})-\frac{35\sqrt{3}}{32}\cos(2\theta_{ij0})-\frac{21\sqrt{3}}{8}\cos(\theta_{ij}+\theta_{ij0})
+\frac{9\sqrt{3}}{64}\cos(\chi-2\theta_{ij})+\frac{129\sqrt{3}}{64}\cos(\chi+2\theta_{ij})\nonumber\\
&&+\frac{15\sqrt{3}}{64}\cos(\chi-2\theta_{ij0})+\frac{9\sqrt{3}}{64}\cos(\chi+2\theta_{ij}-2\theta_{ij0})
+\frac{3\sqrt{3}}{8}\cos(\chi+\theta_{ij}-\theta_{ij0})-\frac{\sqrt{3}}{16}\cos(\chi+3\theta_{ij}-\theta_{ij0})\nonumber\\
&&+\frac{3\sqrt{3}}{16}\cos(\chi-3\theta_{ij}+\theta_{ij0})-\frac{33\sqrt{3}}{8}\cos(\chi-\theta_{ij}+\theta_{ij0})
-\frac{137\sqrt{3}}{64}\cos(\chi+2\theta_{ij0})-\frac{63\sqrt{3}}{64}\cos(\chi-2\theta_{ij}+2\theta_{ij0})\nonumber\\
&&-\frac{9\sqrt{3}}{16}\cos(\chi-\theta_{ij}-\theta_{ij0})+\frac{3\sqrt{3}}{16}\cos(\chi+\theta_{ij}+\theta_{ij0})
+\left(3\sqrt{3}\sin\chi+\frac{27\sqrt{3}}{16}\sin(2\theta_{ij})-3\sqrt{3}\sin(\theta_{ij}-\theta_{ij0})\right.\nonumber\\
&&+3\sqrt{3}\sin(\theta_{ij}+\theta_{ij0})-\frac{21\sqrt{3}}{32}\sin(\chi-2\theta_{ij})+\frac{21\sqrt{3}}{32}\sin(\chi+2\theta_{ij})
+\frac{3\sqrt{3}}{8}\sin(\chi+\theta_{ij}-\theta_{ij0})+\frac{27\sqrt{3}}{8}\sin(\chi\nonumber\\
\label{equ3.53}&&\left.\left.-\theta_{ij}+\theta_{ij0})+\frac{3\sqrt{3}}{8}\sin(\chi-\theta_{ij}-\theta_{ij0})+\frac{27\sqrt{3}}{8}\sin(\chi+\theta_{ij}+\theta_{ij0})\right)\Omega(t-t_{0})\right],
\end{eqnarray}

\begin{eqnarray}
\stackrel{1}{v}_{ij}^{[3]}&=&\frac{\kappa e }{\lambda^{\frac{3}{2}}}R\Omega \left[3\sqrt{3}\sin\chi-6\sqrt{3}\sin(2\theta_{ij})+\frac{\sqrt{3}}{4}\sin(\theta_{ij}-\theta_{ij0})
+\frac{21\sqrt{3}}{16}\sin(2\theta_{ij}-2\theta_{ij0})+\frac{3\sqrt{3}}{8}\sin(3\theta_{ij}-\theta_{ij0})\right.\nonumber\\
&&+\frac{45\sqrt{3}}{8}\sin(\theta_{ij}+\theta_{ij0})-\frac{3\sqrt{3}}{8}\sin(\chi-2\theta_{ij})
-\frac{27\sqrt{3}}{8}\sin(\chi+2\theta_{ij})-\frac{9\sqrt{3}}{32}\sin(\chi+2\theta_{ij}-2\theta_{ij0})
+\frac{3\sqrt{3}}{16}\nonumber\\
&&\times\sin(\chi+3\theta_{ij}-\theta_{ij0})
+\frac{9\sqrt{3}}{16}\sin(\chi-3\theta_{ij}+\theta_{ij0})-\frac{3\sqrt{3}}{4}\sin(\chi-\theta_{ij}+\theta_{ij0})
-\frac{63\sqrt{3}}{32}\sin(\chi-2\theta_{ij}+2\theta_{ij0})\nonumber\\
&&-\frac{3\sqrt{3}}{16}\sin(\chi-\theta_{ij}-\theta_{ij0})+\frac{51\sqrt{3}}{16}\sin(\chi+\theta_{ij}+\theta_{ij0})
+\left(\frac{27\sqrt{3}}{8}\cos(2\theta_{ij})-3\sqrt{3}\cos(\theta_{ij}-\theta_{ij0})+3\sqrt{3}\right.\nonumber\\
&&\times\cos(\theta_{ij}+\theta_{ij0})+\frac{21\sqrt{3}}{16}\cos(\chi-2\theta_{ij})+\frac{21\sqrt{3}}{16}\cos(\chi+2\theta_{ij})
+\frac{3\sqrt{3}}{8}\cos(\chi+\theta_{ij}-\theta_{ij0})-\frac{27\sqrt{3}}{8}\cos(\chi\nonumber\\
\label{equ3.54}&&\left.\left.-\theta_{ij}+\theta_{ij0})-\frac{3\sqrt{3}}{8}\cos(\chi-\theta_{ij}-\theta_{ij0})+\frac{27\sqrt{3}}{8}\cos(\chi+\theta_{ij}+\theta_{ij0})\right)\Omega(t-t_{0})\right],
\end{eqnarray}
where
\begin{equation}\label{equ3.55}
\left\{\begin{array}{l}
\displaystyle \theta_{120}=\Omega t_{0}-\frac{\pi}{3},\smallskip\\
\displaystyle \theta_{230}=\Omega t_{0}-\pi,\smallskip\\
\displaystyle \theta_{310}=\Omega t_{0}-\frac{5\pi}{3}.
\end{array}\right.
\end{equation}
As for the vertex angle $\beta_{ij}$ of Taiji triangle, the angle between the relative radius vectors of SCs $\boldsymbol{r}_{ki},\boldsymbol{r}_{kj}\ (i\neq j)$, according to its definition~(\ref{equ2.32}), its expression to the order of $e^2$ is
\begin{eqnarray}
\label{equ3.56}\beta_{ij}&=&\stackrel{0}{\beta}_{ij}+\stackrel{1}{\beta}_{ij}^{[2]},
\end{eqnarray}
where $\stackrel{0}{\beta}_{ij}$ is certainly the angle between $\stackrel{0}{\boldsymbol{r}}_{ki},\stackrel{0}{\boldsymbol{r}}_{kj}\ (i\neq j)$, and under the optimized case of Eq.~(\ref{equ2.47}), its expression is Eq.~(\ref{equ2.51}). As the cases of $l_{ij}$ and $v_{ij}$, the leading-order correction of $\stackrel{0}{\beta}_{ij}$ due to the attraction of the Earth can also be derived by Eqs.~(\ref{equ3.46})---(\ref{equ3.49}),
\begin{eqnarray}
\stackrel{1}{\beta}_{ij}^{[2]}&=&\frac{\kappa  }{\lambda^{\frac{3}{2}}}\left[\frac{123\sqrt{3}}{64}\cos(2\theta_{ij})-\frac{\sqrt{3}}{16}\cos(3\theta_{ij}-\theta_{ij0})
-\frac{35\sqrt{3}}{64}\cos(2\theta_{ij0})-\frac{21\sqrt{3}}{16}\cos(\theta_{ij}+\theta_{ij0})+\frac{9\sqrt{3}}{128}\cos(\chi-2\theta_{ij})\right.\nonumber\\
&&+\frac{129\sqrt{3}}{128}\cos(\chi+2\theta_{ij})+\frac{15\sqrt{3}}{128}\cos(\chi-2\theta_{ij0})
-\frac{\sqrt{3}}{32}\cos(\chi+3\theta_{ij}-\theta_{ij0})+\frac{3\sqrt{3}}{32}\cos(\chi-3\theta_{ij}+\theta_{ij0})-\frac{137\sqrt{3}}{128}\nonumber\\
&&\times\cos(\chi+2\theta_{ij0})-\frac{9\sqrt{3}}{32}\cos(\chi-\theta_{ij}-\theta_{ij0})
+\frac{3\sqrt{3}}{32}\cos(\chi+\theta_{ij}+\theta_{ij0})
+\left(\frac{27\sqrt{3}}{32}\sin(2\theta_{ij})+\frac{3\sqrt{3}}{2}\sin(\theta_{ij}+\theta_{ij0})\right.\nonumber\\
\label{equ3.57}&&\left.-\frac{21\sqrt{3}}{64}\sin(\chi-2\theta_{ij})+\frac{21\sqrt{3}}{64}\sin(\chi+2\theta_{ij})
+\frac{3\sqrt{3}}{16}\sin(\chi-\theta_{ij}-\theta_{ij0})+\frac{27\sqrt{3}}{16}\sin(\chi+\theta_{ij}+\theta_{ij0})\bigg)\Omega(t-t_{0})\right].
\end{eqnarray}

Eqs.~(\ref{equ2.29}), (\ref{equ3.53})---(\ref{equ3.55}) and (\ref{equ3.57}) show that all the kinematic indicators of Taiji triangle still remain a phase shift of $2\pi/3$, which implies that even under the perturbation of the Earth, three SCs still keep the symmetry in the present orbit model. As shown by Eqs.~(\ref{equ3.3})---(\ref{equ3.6}),
because the semi-major axes of the orbits of the Earth and SCs are the same but the total mass of the Sun and the Earth is different from that of the Sun and each SC, according to Kepler's third law, the average angular velocities of the Earth and SCs are different tinily. In this paper, only Eq.~(\ref{equ3.21}) up to its next-leading-order is focused on, so by Eqs.~(\ref{equ3.12}) and (\ref{equ3.23}), the difference between the average angular velocities of the Earth and SCs does not contribute to the perturbative solutions, and further does not affect the kinematic indicators of Taiji triangle.

As is mentioned before, the barycentre of three SCs also need be corrected by the Earth perturbation. By use of Eqs.~(\ref{equ2.16}) and (\ref{equ3.44}), the radius vector of the barycentre of three SCs perturbed by the Earth is
\begin{equation}\label{equ3.58}
\boldsymbol{r}=\frac{1}{3}(\boldsymbol{r}_{1}+\boldsymbol{r}_{2}+\boldsymbol{r}_{3})
=\stackrel{0}{\boldsymbol{r}}+\stackrel{1}{\boldsymbol{r}}^{[2]}+\stackrel{1}{\boldsymbol{r}}^{[3]},
\end{equation}
where
$\stackrel{0}{\boldsymbol{r}}$ is just $\boldsymbol{r}$ in Eq. (\ref{equ2.21}),
the barycentre of three SCs unperturbed by the Earth, whose three components in heliocentric coordinate system are in Eq.~(\ref{equ2.22}), and $\stackrel{1}{\boldsymbol{r}}^{[2]}$ and $\stackrel{1}{\boldsymbol{r}}^{[3]}$ are the leading-order and next-leading-order corrections to $\stackrel{0}{\boldsymbol{r}}$  due to the attraction of the Earth, respectively. Eq.~(\ref{equ3.29}) shows that $$\stackrel{1}{\boldsymbol{r}}_{1}^{[2]}=\stackrel{1}{\boldsymbol{r}}_{2}^{[2]}
=\stackrel{1}{\boldsymbol{r}}_{3}^{[2]}.$$
Hence,
\begin{equation}\label{equ3.60}
\stackrel{1}{\boldsymbol{r}}^{[2]}=\frac{1}{3}\left(\stackrel{1}{\boldsymbol{r}}_{1}^{[2]}+
\stackrel{1}{\boldsymbol{r}}_{2}^{[2]}+\stackrel{1}{\boldsymbol{r}}_{3}^{[2]}\right)
=\stackrel{1}{\boldsymbol{r}}_{k}^{[2]},\qquad \text{for}\ k=1,2,3.
\end{equation}
Further, by using Eqs.~(\ref{equ3.33}) and (\ref{equ3.34})---(\ref{equ3.43}),
\begin{equation}\label{equ3.61}
\stackrel{1}{\boldsymbol{r}}^{[3]}=\frac{1}{3}\left(\stackrel{1}{\boldsymbol{r}}_{1}^{[3]}
+\stackrel{1}{\boldsymbol{r}}_{2}^{[3]}+\stackrel{1}{\boldsymbol{r}}_{3}^{[3]}\right)
=\frac{\kappa }{\lambda^{\frac{3}{2}}}R\left(E_{xk}'e'+\Lambda_{xk}\lambda^{\frac{3}{2}},E_{yk}'e'
+\Lambda_{yk}\lambda^{\frac{3}{2}},0\right).
\end{equation}
In order to derive the trailing angle of Taiji constellation following the Earth  from the viewpoint of the Sun, the projection of $\boldsymbol{r}$ on the $x$-$y$ plane (the ecliptic plane) needs to be evaluated, and it is
\begin{equation}\label{equ3.62}
\boldsymbol{r}_{\parallel}=\stackrel{0}{\boldsymbol{r}}_{\parallel}+\stackrel{1}{\boldsymbol{r}}^{[2]}
+\stackrel{1}{\boldsymbol{r}}^{[3]}.
\end{equation}
Here, $\stackrel{0}{\boldsymbol{r}}_{\parallel}$ is the projection of $\stackrel{0}{\boldsymbol{r}}$ on the $x$-$y$ plane, and its components are given by the first two expressions in Eq.~(\ref{equ2.22}).  Since
the $z$-components of both $\stackrel{1}{\boldsymbol{r}}^{[2]}$ and $\stackrel{1}{\boldsymbol{r}}^{[3]}$ are zero, the projections of $\stackrel{1}{\boldsymbol{r}}^{[2]}$ and $\stackrel{1}{\boldsymbol{r}}^{[3]}$ on the $x$-$y$ plane are themselves. Thus, the trailing angle of Taiji constellation following the Earth  from the viewpoint of the Sun is defined by
\begin{equation}\label{equ3.63}
\tilde{\chi}=\pm\arccos\left(\frac{\boldsymbol{r}_{\parallel}\cdot \boldsymbol{r}'}{|\boldsymbol{r}_{\parallel}| |\boldsymbol{r}'|}\right),
\end{equation}
where the sign ``$-$'' on the right hand side represents that Taiji constellation is preceding the Earth. With Eqs.~(\ref{equ2.22}), (\ref{equ3.8}), (\ref{equ3.9})---(\ref{equ3.12}) and (\ref{equ3.60})---(\ref{equ3.62}),
the expression of $\tilde{\chi}$ under the optimized case of Eq.~(\ref{equ2.47}) to the order of $e^3$ is
\begin{eqnarray}
\tilde{\chi}&=&\chi+2\sin(\varphi-(\chi+\Omega t))e'-\frac{5}{4}\sin(2\varphi-2(\chi+\Omega t))e'^2+
\left(\frac{13}{12}\sin(3\varphi-3(\chi+\Omega t))-\frac{1}{4}\sin(\varphi-(\chi+\Omega t))\right)e'^3+\Omega t\zeta\nonumber\\
&&+\frac{2}{3}\sin(3\Omega t)e^3+\frac{\kappa }{\lambda^{\frac{3}{2}}}\bigg(-4\sin\chi+2\sin(\Omega t-\Omega t_{0})+\sin(\chi+\Omega t-\Omega t_{0})+3\sin(\chi-\Omega t+\Omega t_{0})+(-2+2\cos\chi)\nonumber\\
&&\times\Omega(t-t_{0})+\frac{3}{2}\sin\chi\Omega^2(t-t_{0})^2\bigg)+\frac{\kappa }{\lambda^{\frac{3}{2}}}\bigg[\bigg(4\sin\chi-\sin(\chi+\Omega t-\Omega t_{0})-3\sin(\chi-\Omega t+\Omega t_{0})-2\cos\chi\Omega(t-t_{0})\nonumber\\
&&\left.-\frac{3}{2}\sin\chi\Omega^2(t-t_{0})^2\right)\lambda^{\frac{3}{2}}+\bigg(\frac{33}{8}\sin(\varphi-\Omega t)
+\frac{39}{4}\sin(\varphi-\chi-\Omega t)+\frac{5}{8}\sin(\varphi-2\chi-\Omega t)-\frac{3}{8}\sin(\varphi+\Omega t-2\Omega t_{0})\nonumber\\
&&-\frac{15}{4}\sin(\varphi-\chi+\Omega t-2\Omega t_{0})-\frac{3}{8}\sin(\varphi-2\chi+\Omega t-2\Omega t_{0})
-\frac{15}{4}\sin(\varphi-\Omega t_{0})-6\sin(\varphi-\chi-\Omega t_{0})-\frac{1}{4}\sin(\varphi\nonumber\\
&&-2\chi-\Omega t_{0})+\Big(\frac{9}{4}\cos(\varphi-\Omega t)+\frac{9}{2}\cos(\varphi-\chi-\Omega t)+\frac{1}{4}\cos(\varphi-2\chi-\Omega t)+\frac{9}{4}\cos(\varphi-\Omega t_{0})+9\cos(\varphi-\chi-\Omega t_{0})\nonumber\\
\label{equ3.64}&&+\frac{3}{4}\cos(\varphi-2\chi-\Omega t_{0})\Big)\Omega (t-t_{0})\bigg)e'\bigg].
\end{eqnarray}

Equation~(\ref{equ3.64}) shows that although the tiny difference between the average angular velocities of the Earth and SCs does not contribute to the kinematic indicators of Taiji triangle, but it can correct the trailing angle of Taiji constellation at the order of $e'^3(e^3)$. Moreover, since the difference between the average angular velocities of the Earth and SCs is very tiny, the Earth perturbs SCs in an almost resonance way, which leads to the existence of the terms characterized by $\Omega(t-t_{0})$ and $\Omega^2(t-t_{0})^2$ in the results related to the perturbation of the Earth so that these results will grow unboundedly over time.

As indicated by Eqs.~(\ref{equ3.50}), (\ref{equ3.51}) and (\ref{equ3.56}), $\stackrel{1}{l}_{ij}^{[3]}$ and $\stackrel{1}{v}_{ij}^{[3]}$ are the corrections to $\stackrel{0}{l}_{ij}$ and $\stackrel{0}{v}_{ij}$ at the order of $e^3$, respectively, and $\stackrel{1}{\beta}_{ij}^{[2]}$
is the correction to $\stackrel{0}{\beta}_{ij}$ at the order of $e^2$, so the following conditions must be hold:
\begin{equation}\label{equ3.65}
\stackrel{1}{l}_{ij}^{[3]}\ll\stackrel{0}{l}_{ij}^{[2]},\qquad \stackrel{1}{v}_{ij}^{[3]}\ll\stackrel{0}{v}_{ij}^{[2]},\qquad
\stackrel{1}{\beta}_{ij}^{[2]}\ll\stackrel{0}{\beta}_{ij}^{[1]}.
\end{equation}
When $\Omega(t-t_{0})>1$, Eqs.~(\ref{equ3.53}), (\ref{equ3.54}) and (\ref{equ3.57}) show that the magnitudes of $\stackrel{1}{l}_{ij}^{[3]},\stackrel{1}{v}_{ij}^{[3]}$
and $\stackrel{1}{\beta}_{ij}^{[2]}$ are determined by $\frac{\kappa e }{\lambda^{3/2}}R\Omega(t-t_{0}), \frac{\kappa e }{\lambda^{3/2}}R\Omega^2(t-t_{0})$ and $\frac{\kappa}{\lambda^{3/2}}\Omega(t-t_{0})$, respectively, and by  Eqs.~(\ref{equ2.49})---(\ref{equ2.51}), those of $\stackrel{0}{l}_{ij}^{[2]},\stackrel{0}{v}_{ij}^{[2]}$ and
$\stackrel{0}{\beta}_{ij}^{[1]}$ are determined by $Re^2,R\Omega e^2$ and $e$. Thus, Eq.~(\ref{equ3.65}) brings about
\begin{equation}\label{equ3.66}
\Omega(t-t_{0})\ll \frac{e}{\kappa/\lambda^{3/2}}\approx79.76.
\end{equation}
This is just a very loose constraint on the running period of Taiji program.
In fact, in order that the perturbation expansion is valid, the conditions
\begin{equation}\label{equ3.67}
\big|\stackrel{1}{\boldsymbol{r}}_{k}^{[2]}\big|\ll\big|\stackrel{0}{\boldsymbol{r}}_{k}^{[1]}\big|,\qquad \big|\stackrel{1}{\boldsymbol{r}}_{k}^{[3]}\big|\ll\big|\stackrel{0}{\boldsymbol{r}}_{k}^{[2]}\big|
\end{equation}
are required. Similarly to the above discussion, when $\Omega(t-t_{0})>1$, Eqs.~(\ref{equ3.29}), (\ref{equ3.33}) and (\ref{equ3.34})---(\ref{equ3.43}) show that the magnitudes of $\big|\stackrel{1}{\boldsymbol{r}}_{k}^{[2]}\big|$ and
$\big|\stackrel{1}{\boldsymbol{r}}_{k}^{[3]}\big|$ are determined by $\frac{\kappa}{\lambda^{3/2}}R\Omega^2(t-t_{0})^2$ and $\frac{\kappa e }{\lambda^{3/2}}R\Omega^2(t-t_{0})^2$, respectively, and by Eqs.~(\ref{equ2.16}) and (\ref{equ2.17})---(\ref{equ2.20}), those of $\big|\stackrel{0}{\boldsymbol{r}}_{k}^{[1]}\big|$ and
$\big|\stackrel{0}{\boldsymbol{r}}_{k}^{[2]}\big|$ are determined by $Re$ and $Re^2$, which gives
\begin{equation}\label{equ3.68}
\Omega(t-t_{0})\ll \left(\frac{e}{\kappa/\lambda^{3/2}}\right)^{\frac{1}{2}}\approx8.931.
\end{equation}
From the above inequality, the running period of Taiji program should be set reasonably to be $8.931/3\approx3$ years, so that Taiji configuration needs to be restored to their initial state every $3$ years if one wishes to lengthen the lifetime of Taiji program.  It should be pointed out that the same conclusion can also be drawn by the expression~(\ref{equ3.64}) of the trailing angle of Taiji constellation following the Earth  from the viewpoint of the Sun.
\section{Summary and discussions~\label{Sec:fourth}}
The ground-based detectors are unable to detect GWs below $0.1$ Hz~\cite{Danzmann:1997hm,Harms:2013raa}, so the space-based GW detector, like LISA~\cite{Dhurandhar:2004rv,Nayak:2006zm}, will become increasingly important.
Taiji program~\cite{Hu:2017mde,Wu:2018clg}, set up by the Chinese Academy of Sciences, is composed of three identical SCs, and these SCs orbit the Sun and form a triangle whose nominal side is about $3\times10^6$ km. Taiji program will
observe GWs covering the ranges from $0.1$ mHz to $1.0$ Hz by using coherent laser beams exchanged between three SCs.
In this paper, the model in Refs.~\cite{Dhurandhar:2004rv,Nayak:2006zm} designed for LISA mission could be used in the Taiji program with the replacement of the orbit parameters due to the similarity between Taiji program and LISA mission, and the orbits of Taiji SCs are analytically analyzed in the heliocentric coordinate system in a higher precision than those of LISA SCs~\cite{Dhurandhar:2004rv,Nayak:2006zm,Dhurandhar:2008yu,Pucacco:2010mn} in the Hill system or CW system~\cite{Pucacco:2010mn}.

As the primary task, the unperturbed Keplerian orbits of Taiji SCs are expanded to $e^3$ order firstly. Then, the three arm-lengths of Taiji triangle and their rates of change are also expanded to $e^3$ order, while the three vertex angles of the triangle are expanded to $e^2$ order. The expansions of these kinematic indicators of Taiji triangle show that only under the cases of the angle $\phi=\pm\pi/3$, Taiji triangle is approximately equilateral one.
However even so, the relative motion between SCs still exists, which results in that the first generation TDI does not work so that modified first generation TDI or further, the second generation TDI needs to be used~\cite{Dhurandhar:2008yu,Tinto:2003vj,Vallisneri:2005ji,Tinto:2014lxa}. Moreover, the relative motion between SCs also causes the Doppler shift of the laser frequency, which will interfere with the measurement of GWs~\cite{Nayak:2006zm}. In order to consider a simpler TDI strategy as LISA in Refs.~\cite{Nayak:2006zm,Dhurandhar:2008yu} and lower the adverse effect of Doppler shift of the laser frequency, it has been shown that by adjusting the angle $\phi$ around $\pm\pi/3$ at $e^1$ order, the variations of all the kinematic indicators of Taiji triangle are minimized, respectively, which is compatible to that of LISA in Refs.~\cite{Nayak:2006zm,DeMarchi:2011ye}.
Further, all the optimized expressions of these kinematic indicators of Taiji triangle are presented in this paper.

The implementation of TDI~\cite{Pucacco:2010mn} requires the relative motion between SCs analyzed more accurately, which means that the perturbations of the Earth, the Moon, and other planets on SCs should be considered. In the present paper, only the perturbation of the Earth on SCs is focused on, as did for LISA in Ref.~\cite{Dhurandhar:2008yu}, and the precision of the orbits is enhanced to the order of $e^3$. In the ecliptic plane, the $x$ axis of the heliocentric coordinate system may be chosen arbitrarily relative to the major axis of the Earth's orbit, and then, the nominal trailing angle of the Taiji constellation following the Earth is set to be $\chi(\approx\pm\pi/9)$ from the viewpoint of the Sun by adjusting the mean anomaly of the Earth Kepler's equation, where the negative value of $\chi$ means that the constellation is preceding the Earth. Under the theoretical framework of the problem of three bodies~\cite{moulton1960}, the leading-order and the next-leading-order perturbative solutions are derived, and it has been shown that the former do not contribute to the variations of the kinematic indicators of Taiji triangle, but the latter do. Moreover, both of them can lead to the change of the above trailing angle of Taiji constellation. Compared with the previous discussion on the Earth perturbation on LISA SCs~\cite{Dhurandhar:2008yu,Pucacco:2010mn}, our results include the influence of the interaction between the Sun and the Earth and the effect of the eccentricity $e'$ of the Earth's orbit.

Further, the influence of the Earth perturbation on Taiji configuration is discussed. The analytic expressions for the
leading-order corrections to all the kinematic indicators of Taiji triangle resulted from the Earth and the modified expression for the above trailing angle of Taiji constellation to the order of $e^3$ are first presented in this paper. Since the difference between the average angular velocities of the Earth and SCs is very tiny, it only contributes to
the corrected trailing angle of Taiji constellation. Moreover, it is this tiny difference which leads to the Earth perturbing SCs in an almost resonance way, so that the terms characterized by $\Omega(t-t_{0})$ and $\Omega^2(t-t_{0})^2$ exist in the results related to the perturbation of the Earth, and then, these results will grow unboundedly over time, as indicated by Ref.~\cite{Dhurandhar:2008yu}. In order to avoid this case, by the results in this paper, the running period of Taiji should be set reasonably to be about $3$ years, and Taiji configuration needs to be restored to their initial state every $3$ years if one wishes to lengthen its running time.

As far as we know, the results up to the order of $e^3$, and especially those related to the perturbation of the Earth on SCs have not been given before for Taiji or LISA, so the results in the present paper may be useful for their development. As mentioned before, LISA and Taiji might be in operation at the same time for a period.
According to the results in this paper, both of $\varphi$ and $\chi$ have two choices, so in terms of LISA and Taiji,
there are actually four kinds of combinations to be chosen. No matter which combination is chosen, their simultaneous operation will improve the sensitivity and angle resolution of detecting GWs, which will prompt the further study on GW significantly. Having discussed the perturbation of the Earth, the next task is further to take into account the perturbations of the other celestial bodies. Among them, the most important one is the Jupiter. As indicated in Ref.~\cite{Dhurandhar:2008yu}, the leading-order perturbative effect of the Jupiter is less than 10\% than that of the Earth, and because the Jupiter perturbing SCs is not in resonance, its perturbative effect will not accumulate in the first few years. In view of this, the perturbative solutions in this paper may be extended to include the perturbative effect of the Jupiter. Moreover, the perturbations of the Moon and the other planets also need to be analyzed carefully in the following task.

\begin{acknowledgements}
This work was supported, in part, by the Strategic Priority Research Program of the Chinese Academy of Sciences, Grants No. XDB23030100 and No. XDB23040000, by the National Natural Science Foundation of China (NSFC) under Grants No.~11690022 and No. 11635009, and by the Ministry of Science and Technology of the People's Republic of China
(2015CB856703).
\end{acknowledgements}

\appendix\label{appendix}
\section{DERIVATION \uppercase{of} Eq.~(\ref{equ2.16})}
As mentioned before, the Kepler's equations (\ref{equ2.2}), (\ref{equ2.7}), and (\ref{equ2.8}) need to be dealt with by the iterative method under the case of the eccentricity $e \ll 1$. Then, the expression of $\psi_{k}$ expanded to $e^3$ order is
\begin{widetext}
\begin{eqnarray}
\label{equ2.10}\psi_{k}&=&\sigma_{k}-(\sin\sigma_{k})e+\frac{\sin(2\sigma_{k})}{2}e^2
+\left(\frac{1}{8}\sin\sigma_{k}-\frac{3}{8}\sin(3\sigma_{k})\right)e^3,
\end{eqnarray}
where
\begin{equation}\label{equ2.11}
\sigma_{k}=\Omega t-(k-1)\frac{2\pi}{3},\qquad k=1,2,3,
\end{equation}
and further,
\begin{eqnarray}
\label{equ2.12}\cos\psi_{k}&=&\cos\sigma_{k}+\left(\frac{1}{2}-\frac{1}{2}\cos(2\sigma_{k})\right)e
+\left(-\frac{3}{8}\cos\sigma_{k}+\frac{3}{8}\cos(3\sigma_{k})\right)e^2
+\left(\frac{1}{3}\cos(2\sigma_{k})-\frac{1}{3}\cos(4\sigma_{k})\right)e^3,\\
\label{equ2.13}\sin\psi_{k}&=&\sin\sigma_{k}
-\frac{1}{2}\sin(2\sigma_{k})e
+\left(-\frac{1}{8}\sin\sigma_{k}+\frac{3}{8}\sin(3\sigma_{k})\right)e^2
+\left(\frac{1}{6}\sin(2\sigma_{k})-\frac{1}{3}\sin(4\sigma_{k})\right)e^3.
\end{eqnarray}
As for the parameter $\alpha$, by Eq. (\ref{equ2.4}),
\begin{equation}\label{equ2.14}
\alpha=\frac{\sqrt{3}}{2\cos\phi}e-\frac{\sqrt{3}}{4}\frac{\tan^2\phi}{\cos\phi}e^2+\frac{\sqrt{3}}{4}\frac{\tan^2\phi}{\cos^3\phi}e^3,
\end{equation}
and with it,  Eq. (\ref{equ2.3}) gives
\begin{equation}\label{equ2.15}
\left\{\begin{array}{l}
\displaystyle \cos\varepsilon=1-\frac{1}{2}(\tan^2\phi)e^2+\left(\tan^2\phi+\frac{1}{2}\tan^4\phi\right)e^3,\smallskip\\
\displaystyle \sin\varepsilon=\left(\tan\phi\right) e+\left(-\tan\phi-\frac{1}{2}\tan^3\phi\right)e^2
+\left(\frac{\tan\phi}{\cos^2\phi}+\frac{1}{2}\tan^5\phi\right)e^3.
\end{array}\right.
\end{equation}

With all the above results, the combination of Eqs.~(\ref{equ2.1}) and (\ref{equ2.3})---(\ref{equ2.6}) can bring about the expansion of $\boldsymbol{r}_{k}\ (k=1,2,3)$ to $e^3$ order, namely Eq.~(\ref{equ2.16}), with
\begin{equation}\label{equ2.17}
\left\{\begin{array}{l}
\displaystyle x_{k}^{(0)}=R\cos(\Omega t),\smallskip\\
\displaystyle y_{k}^{(0)}=R\sin(\Omega t),\smallskip\\
\displaystyle z_{k}^{(0)}=0,
\end{array}\right.
\end{equation}
\begin{equation}\label{equ2.18}
\left\{\begin{array}{l}
\displaystyle x_{k}^{(1)}=R\left[\frac{3}{2}\cos\left((k-1)\frac{2\pi}{3}\right)-\frac{1}{2}\cos\left(2\sigma_{k}+(k-1)\frac{2\pi}{3}\right)\right],\smallskip\\
\displaystyle y_{k}^{(1)}=R\left[3\sqrt{3}(k-1)\cos^k\left(\frac{2\pi}{3}\right)-\frac{1}{2}\sin\left(2\sigma_{k}+(k-1)\frac{2\pi}{3}\right)\right],\smallskip\\
\displaystyle z_{k}^{(1)}=R\tan\phi \cos\sigma_{k},
\end{array}\right.
\end{equation}
\begin{equation}\label{equ2.19}
\left\{\begin{array}{l}
\displaystyle x_{k}^{(2)}=R\left[\left(\frac{1}{8}-\frac{1}{4}\tan^2\phi\right)\cos\left(\sigma_{k}-(k-1)\frac{2\pi}{3}\right)
-\left(\frac{1}{2}+\frac{1}{4}\tan^2\phi\right)\cos\left(\sigma_{k}+(k-1)\frac{2\pi}{3}\right)
+\frac{3}{8}\cos\left(3\sigma_{k}+(k-1)\frac{2\pi}{3}\right)\right],\smallskip\\
\displaystyle y_{k}^{(2)}=R\left[\left(\frac{\sqrt{3}}{2}-\sqrt{3}\tan^2\phi\right)(k-1)\cos^k\left(\frac{2\pi}{3}\right)\cos\sigma_{k}
-\frac{5}{8}\sin\left(\sigma_{k}+(k-1)\frac{2\pi}{3}\right)
+\frac{3}{8}\sin\left(3\sigma_{k}+(k-1)\frac{2\pi}{3}\right)\right],\smallskip\\
\displaystyle z_{k}^{(2)}=R\tan\phi\left[\frac{3}{2}-\left(1+\frac{1}{2}\tan^2\phi\right)\cos\sigma_{k}
-\frac{1}{2}\cos\left(2\sigma_{k}\right)\right],
\end{array}\right.
\end{equation}
\begin{equation}\label{equ2.20}
\left\{\begin{array}{l}
\displaystyle x_{k}^{(3)}=R\left[\cos\left((k-1)\frac{2\pi}{3}\right)\left(-\frac{3}{4}\tan^2\phi+\left(1+\frac{1}{2}\tan^2\phi\right)\tan^2\phi \cos\sigma_{k}
+\frac{1}{4}\tan^2\phi\cos(2\sigma_{k})\right)-\frac{1}{24}\cos\left(2\sigma_{k}-(k-1)\frac{2\pi}{3}\right)\right.\smallskip\\
\displaystyle\phantom{x_{2}^{(3)}=}\left.+\frac{3}{8}\cos\left(2\sigma_{k}+(k-1)\frac{2\pi}{3}\right)-\frac{1}{3}\cos\left(4\sigma_{k}+(k-1)\frac{2\pi}{3}\right)\right],\smallskip\\
\displaystyle y_{k}^{(3)}=R\left[(k-1)\cos^k\left(\frac{2\pi}{3}\right)\left(-\frac{3\sqrt{3}}{2}\tan^2\phi+\left(2\sqrt{3}+\sqrt{3}\tan^2\phi\right)\tan^2\phi\cos\sigma_{k}
+\left(\frac{3\sqrt{3}}{2}+\frac{\sqrt{3}}{2}\tan^2\phi\right)\cos(2\sigma_{k})\right)\right.\smallskip\\
\displaystyle\phantom{x_{2}^{(3)}=}\left.+\frac{5}{12}\sin\left(2\sigma_{k}-(k-1)\frac{2\pi}{3}\right)-\frac{1}{3}\sin\left(4\sigma_{k}+(k-1)\frac{2\pi}{3}\right)\right],\smallskip\\
\displaystyle z_{k}^{(3)}=R\tan\phi\left[-\frac{3}{2}-\frac{3}{4}\tan^2\phi+\left(\frac{5}{8}+\tan^2\phi+\frac{1}{2}\tan^4\phi\right)\cos\sigma_{k}+\left(\frac{1}{2}+\frac{1}{4}\tan^2\phi\right)\cos(2\sigma_{k})+\frac{3}{8}\cos(3\sigma_{k})\right].\smallskip\\
\end{array}\right.
\end{equation}

\section{DERIVATION \uppercase{of} Eq.~(\ref{equ3.8})}
Because $e'$ and $e$ have the same order of magnitudes, like the cases of Eqs.~(\ref{equ2.2}), (\ref{equ2.7}) and (\ref{equ2.8}), $\psi'$ in the Kepler's equation (\ref{equ3.2}) can be and only needs to be expanded to $e'^3$ order. Moreover, Eq.~(\ref{equ3.6}) shows that the order of $\zeta$ is the same as that of $e'^3$, so $\psi'$ also needs to be expanded to $\zeta^1$ order,
\begin{eqnarray}
\label{equ3.7}\psi'&=&\Omega t+\chi-\varphi-\sin\left(\Omega t+\chi-\varphi\right)e'+\frac{\sin\big(2\left(\Omega t+\chi-\varphi\right)\big)}{2}e'^2\nonumber\\
&&+\left(\frac{1}{8}\sin\left(\Omega t+\chi-\varphi\right)-\frac{3}{8}\sin\big(3\left(\Omega t+\chi-\varphi\right)\big)\right)e'^3+(\Omega t)\zeta.
\end{eqnarray}
Applying it to Eq.~(\ref{equ3.1}), the expansion of $\boldsymbol{r}'$ to $e'^3$ order and $\zeta^1$ order is the Eq.~(\ref{equ3.8}) with
\begin{equation}\label{equ3.9}
\left\{\begin{array}{l}
\displaystyle x'^{[0]}=R\cos(\chi+\Omega t),\smallskip\\
\displaystyle y'^{[0]}=R\sin(\chi+\Omega t),\smallskip\\
\displaystyle z'^{[0]}=0,
\end{array}\right.
\end{equation}
\begin{equation}\label{equ3.10}
\left\{\begin{array}{l}
\displaystyle x'^{[1]}=Re'\left(\frac{3}{2}\cos\varphi-\frac{1}{2}\cos\left(\varphi-2(\chi+\Omega t)\right)\right),\smallskip\\
\displaystyle y'^{[1]}=Re'\left(\frac{3}{2}\sin\varphi+\frac{1}{2}\sin\left(\varphi-2(\chi+\Omega t)\right)\right),\smallskip\\
\displaystyle z'^{[1]}=0,
\end{array}\right.
\end{equation}
\begin{equation}\label{equ3.11}
\left\{\begin{array}{l}
\displaystyle x'^{[2]}=Re'^2\left(\frac{3}{8}\cos(2\varphi-3(\chi+\Omega t))+\frac{1}{8}\cos(2\varphi-(\chi+\Omega t))-\frac{1}{2}\cos(\chi+\Omega t)\right),\smallskip\\
\displaystyle y'^{[2]}=Re'^2\left(-\frac{3}{8}\sin(2\varphi-3(\chi+\Omega t))+\frac{1}{8}\sin(2\varphi-(\chi+\Omega t))-\frac{1}{2}\sin(\chi+\Omega t)\right),\smallskip\\
\displaystyle z'^{[2]}=0,
\end{array}\right.
\end{equation}
\begin{equation}\label{equ3.12}
\left\{\begin{array}{l}
\displaystyle x'^{[3]}=R\left[e'^3\left(-\frac{1}{3}\cos(3\varphi-4(\chi+\Omega t))+\frac{3}{8}\cos(\varphi-2(\chi+\Omega t))-\frac{1}{24}\cos(3\varphi-2(\chi+\Omega t))\right)-\zeta \Omega t\sin(\chi+\Omega t)\right],\smallskip\\
\displaystyle y'^{[3]}=R\left[e'^3\left(\frac{1}{3}\sin(3\varphi-4(\chi+\Omega t))-\frac{3}{8}\sin(\varphi-2(\chi+\Omega t))-\frac{1}{24}\sin(3\varphi-2(\chi+\Omega t))\right)+\zeta \Omega t\cos(\chi+\Omega t)\right],\smallskip\\
\displaystyle z'^{[3]}=0.
\end{array}\right.
\end{equation}

\section{DERIVATION \uppercase{of} Eq.~(\ref{equ3.20}) and Solving Equations of $\stackrel{1}{\boldsymbol{r}}_{k}^{[2]},\stackrel{1}{\boldsymbol{r}}_{k}^{[3]}$}
As indicated by Eq.~(\ref{equ3.17}), $\boldsymbol{r}_{k}=\stackrel{0}{\boldsymbol{r}}_{k}+\stackrel{1}{\boldsymbol{r}}_{k}$.
Here, $\stackrel{0}{\boldsymbol{r}}_{k}$, as the unperturbed Keplerian orbit of SC$k$, satisfies the equation of SC$k$ only attracted by the Sun:
\begin{equation}\label{equ3.18}
\frac{d^2\stackrel{0}{\boldsymbol{r}}_{k}}{dt^2}+\frac{\mu\stackrel{0}{\boldsymbol{r}}_{k}}{\big(\stackrel{0}{r}_{k}\big)^3}=0.
\end{equation}
$\stackrel{0}{\boldsymbol{r}}_{k}$ is exactly the solution of the homogeneous equation of Eq.~(\ref{equ3.16}). Therefore, by applying Eqs.~(\ref{equ2.1}), (\ref{equ2.5}) and (\ref{equ2.6}) to Eq.~(\ref{equ3.18}), there is
\begin{equation}\label{equ3.19}
\mu=R^3\Omega^2,
\end{equation}
which is just the Kepler's third law. Inserting Eqs.~(\ref{equ3.17}) and (\ref{equ3.18}) into Eq.~(\ref{equ3.16}) gives
the equation of the perturbative solution $\stackrel{1}{\boldsymbol{r}}_{k}$, and that is Eq.~(\ref{equ3.20}).
With Eq.~(\ref{equ3.19}), Eq.~(\ref{equ3.20}) can be rewritten in the dimensionless form:
\begin{equation}\label{equ3.21}
\frac{1}{\Omega^2}\frac{d^2}{dt^2}\left(\frac{\stackrel{1}{\boldsymbol{r}}_{k}}{R}\right)
+\frac{R^3}{\big(\stackrel{0}{r}_{k}\big)^3}\frac{\stackrel{1}{\boldsymbol{r}}_{k}}{R}
-\frac{3R^5}{\big(\stackrel{0}{r}_{k}\big)^5}\left(\frac{\stackrel{0}{\boldsymbol{r}}_{k}}{R}\cdot\frac{\stackrel{1}{\boldsymbol{r}}_{k}}{R}\right)
\frac{\stackrel{0}{\boldsymbol{r}}_{k}}{R}=
\kappa\left(\frac{R^2\left(\boldsymbol{r}'-\stackrel{0}{\boldsymbol{r}}_{k}\right)}{\big|\boldsymbol{r}'-\stackrel{0}{\boldsymbol{r}}_{k}\big|^3}-\frac{R^2\boldsymbol{r}'}{r'^3}\right).
\end{equation}
By Eqs.~(\ref{equ2.1}), (\ref{equ2.5}), and (\ref{equ2.6}),
\begin{equation}\label{equ2.9}
  \stackrel{0}{r}_{k}=|\stackrel{0}{\boldsymbol{r}}_{k}|=R(1+e\cos{\psi_{k}}),
\end{equation}
where $\stackrel{0}{\boldsymbol{r}}_{k}=\stackrel{0}{\boldsymbol{r}}_{k}^{[0]}
+\stackrel{0}{\boldsymbol{r}}_{k}^{[1]}+\cdots$ according to Eq.~(\ref{equ2.16}),
and then, form Eq.~(\ref{equ2.9}), there are
\begin{eqnarray}\label{equ3.22}
\frac{R^3}{\big(\stackrel{0}{r}_{k}\big)^3}=1-3\left(\cos\sigma_{k}\right)e+\cdots,\qquad
\frac{R^5}{\big(\stackrel{0}{r}_{k}\big)^5}=1-5\left(\cos\sigma_{k}\right)e+\cdots,
\end{eqnarray}
and moreover, by Eq.~(\ref{equ3.8}),
\begin{equation}\label{equ3.23}
\boldsymbol{r}'=\boldsymbol{r}'^{[0]}+\boldsymbol{r}'^{[1]}+\cdots.
\end{equation}
Remember that the perturbative solution $\stackrel{1}{\boldsymbol{r}}_{k}$ is assumed to be expanded in terms of the order of $e$, 
as shown by Eq.~(\ref{equ3.24}).
Now, Eq.~(\ref{equ3.21}) can be expanded in terms of the order of $e$ by using Eqs.~(\ref{equ3.22}), (\ref{equ3.23}), and (\ref{equ3.24}). The leading-order of Eq.~(\ref{equ3.21}) reads
\begin{equation}\label{equ3.25}
\frac{1}{\Omega^2}\frac{d^2}{dt^2}\left(\frac{\stackrel{1}{\boldsymbol{r}}_{k}^{[2]}}{R}\right)
+\frac{\stackrel{1}{\boldsymbol{r}}_{k}^{[2]}}{R}
-3\left(\frac{\stackrel{0}{\boldsymbol{r}}_{k}^{[0]}}{R}\cdot
\frac{\stackrel{1}{\boldsymbol{r}}_{k}^{[2]}}{R}\right)
\frac{\stackrel{0}{\boldsymbol{r}}_{k}^{[0]}}{R}=
\frac{\kappa }{\lambda^{\frac{3}{2}}}\frac{\boldsymbol{r}'^{[0]}-\stackrel{0}{\boldsymbol{r}}_{k}^{[0]}}{R},
\end{equation}
where the second term on the right hand side of Eq.~(\ref{equ3.21}), the interaction between the Sun and the Earth, has been omitted because it is suppressed by $\lambda^{\frac{3}{2}}\approx4.189\times10^{-2}$. According to the
initial condition~(\ref{equ3.27}) of Eq.~(\ref{equ3.20}), the initial condition of Eq.~(\ref{equ3.25}) should be
\begin{equation}\label{equ3.28}
\stackrel{1}{\boldsymbol{r}}_{k}^{[2]}|_{t=t_0}=0,\qquad \frac{d\stackrel{1}{\boldsymbol{r}}_{k}^{[2]}}{dt}\Big|_{t=t_0}=0.
\end{equation}
Thus, the solution of Eq.~(\ref{equ3.25}), under the initial condition (\ref{equ3.28}), is
$\stackrel{1}{\boldsymbol{r}}_{k}^{[2]}
=\Big(\stackrel{1}{x}_{k}^{[2]},\stackrel{1}{y}_{k}^{[2]},\stackrel{1}{z}_{k}^{[2]}\Big)$  with
\begin{equation}\label{equ3.29}
\left\{\begin{array}{l}
\displaystyle \stackrel{1}{x}_{k}^{[2]}=\frac{\kappa }{\lambda^{\frac{3}{2}}}R\left[-\cos(\Omega t)+\frac{3}{2}\cos(\Omega t_{0})-\frac{3}{2}\cos(\chi-\Omega t)+\frac{5}{2}\cos(\chi+\Omega t)+\frac{3}{4}\cos(\chi-\Omega t_{0})-\frac{1}{2}\cos(2\Omega t-\Omega t_{0})\right.\smallskip\\
\displaystyle\phantom{ \big(x_{k}^{\{1\}}\big)^{[0]}=}-\frac{1}{4}\cos(\chi+2\Omega t-\Omega t_{0})-\frac{9}{4}\cos(\chi+\Omega t_{0})+\frac{3}{4}\cos(\chi-2\Omega t+\Omega t_{0})+\left(-2\sin(\Omega t)+2\sin(\chi+\Omega t)\right)\Omega(t-t_{0})\smallskip\\
\displaystyle\phantom{ \big(x_{k}^{\{1\}}\big)^{[0]}=}\left.+\left(\frac{3}{4}\cos(\chi-\Omega t)-\frac{3}{4}\cos(\chi+\Omega t)\right)\Omega^2(t-t_{0})^2\right],\smallskip\\
\displaystyle  \stackrel{1}{y}_{k}^{[2]}=\frac{\kappa }{\lambda^{\frac{3}{2}}}R\left[-\sin(\Omega t)+\frac{3}{2}\sin(\Omega t_{0})+\frac{3}{2}\sin(\chi-\Omega t)+\frac{5}{2}\sin(\chi+\Omega t)-\frac{3}{4}\sin(\chi-\Omega t_{0})-\frac{1}{2}\sin(2\Omega t-\Omega t_{0})\right.\smallskip\\
\displaystyle\phantom{ \big(x_{k}^{\{1\}}\big)^{[0]}=}-\frac{1}{4}\sin(\chi+2\Omega t-\Omega t_{0})-\frac{9}{4}\sin(\chi+\Omega t_{0})-\frac{3}{4}\sin(\chi-2\Omega t+\Omega t_{0})+\left(2\cos(\Omega t)-2\cos(\chi+\Omega t)\right)\Omega(t-t_{0})\smallskip\\
\displaystyle\phantom{ \big(x_{k}^{\{1\}}\big)^{[0]}=}\left.+\left(-\frac{3}{4}\sin(\chi-\Omega t)-\frac{3}{4}\sin(\chi+\Omega t)\right)\Omega^2(t-t_{0})^2\right],\smallskip\\
\displaystyle  \stackrel{1}{z}_{k}^{[2]}=0.
\end{array}\right.
\end{equation}
The next-leading-order of Eq.~(\ref{equ3.21}) can be obtained with the help of Eq.~(\ref{equ3.25}),
\begin{equation}\label{equ3.30}
\frac{1}{\Omega^2}\frac{d^2}{dt^2}\left(\frac{\stackrel{1}{\boldsymbol{r}}_{k}^{[3]}}{R}\right)
+\frac{\stackrel{1}{\boldsymbol{r}}_{k}^{[3]}}{R}
-3\left(\frac{\stackrel{0}{\boldsymbol{r}}_{k}^{[0]}}{R}\cdot\frac{\stackrel{1}{\boldsymbol{r}}_{k}^{[3]}}{R}\right)
\frac{\stackrel{0}{\boldsymbol{r}}_{k}^{[0]}}{R}=
\frac{\kappa }{\lambda^{\frac{3}{2}}}\boldsymbol{D}_{k}-\boldsymbol{A}_{k},
\end{equation}
where
\begin{equation}\label{equ3.31}
\left\{\begin{array}{l}
\displaystyle \boldsymbol{A}_{k}:=-3e\cos\sigma_{k}\frac{\stackrel{1}{\boldsymbol{r}}_{k}^{[2]}}{R}-3\left[(-5e\cos\sigma_{k})
\left(\frac{\stackrel{0}{\boldsymbol{r}}_{k}^{[0]}}{R}\cdot\frac{\stackrel{1}{\boldsymbol{r}}_{k}^{[2]}}{R}\right)
\frac{\stackrel{0}{\boldsymbol{r}}_{k}^{[0]}}{R}+\left(\frac{\stackrel{0}{\boldsymbol{r}}_{k}^{[1]}}{R}\cdot
\frac{\stackrel{1}{\boldsymbol{r}}_{k}^{[2]}}{R}\right)\frac{\stackrel{0}{\boldsymbol{r}}_{k}^{[0]}}{R}\right .\smallskip\\
\displaystyle\phantom{\boldsymbol{A}_{k}:=}\left . +\left(\frac{\stackrel{0}{\boldsymbol{r}}_{k}^{[0]}}{R}\cdot\frac{\stackrel{1}{\boldsymbol{r}}_{k}^{[2]}}{R}\right)
\frac{\stackrel{0}{\boldsymbol{r}}_{k}^{[1]}}{R}\right],\smallskip\\
\displaystyle \boldsymbol{D}_{k}:=\frac{\boldsymbol{r}'^{[1]}-\stackrel{0}{\boldsymbol{r}}_{k}^{[1]}}{R}-\frac{3}{\lambda}
\left[\left(\frac{\boldsymbol{r}'^{[0]}-\stackrel{0}{\boldsymbol{r}}_{k}^{[0]}}{R}\cdot
\frac{\boldsymbol{r}'^{[1]}-\stackrel{0}{\boldsymbol{r}}_{k}^{[1]}}{R}\right)
\frac{\boldsymbol{r}'^{[0]}-\stackrel{0}{\boldsymbol{r}}_{k}^{[0]}}{R}\right]-\lambda^{\frac{3}{2}}\frac{\boldsymbol{r}'^{[0]}}{R}.
\end{array}\right.
\end{equation}
The fact that the expression of $\boldsymbol{A}_{k}$ depends on $\stackrel{1}{\boldsymbol{r}}_{k}^{[2]}$ means that the solution of Eq.~(\ref{equ3.30}), as the next-leading-order perturbative solution, is dependent on the leading-order perturbative solution. The last term of $\boldsymbol{D}_{k}$ shows that the interaction between the Sun and the Earth, from the second term on the right hand side of Eq.~(\ref{equ3.21}), has been considered in Eq.~(\ref{equ3.30}).
From Eq.~(\ref{equ3.27}), the initial condition of Eq.~(\ref{equ3.30}) should be
\begin{equation}\label{equ3.32}
\stackrel{1}{\boldsymbol{r}}_{k}^{[3]}|_{t=t_0}=0,\qquad \frac{d\stackrel{1}{\boldsymbol{r}}_{k}^{[3]}}{dt}\Big|_{t=t_0}=0,
\end{equation}
and then, by inserting the perturbative solution (\ref{equ3.29}) of the leading order into $\boldsymbol{A}_{k}$, the solution of Eq.~(\ref{equ3.30}) is
$\stackrel{1}{\boldsymbol{r}}_{k}^{[3]}=({\stackrel{1}{x}_{k}^{[3]},} {\stackrel{1}{y}_{k}^{[3]},} {\stackrel{1}{z}_{k}^{[3]}})$ presented by Eq.~(\ref{equ3.33}) with
\begin{eqnarray}
E_{xk}&=&-\frac{37}{16}\cos(2\rho_k)+\frac{45}{32}\cos(\chi-2\rho_k)-\frac{111}{32}\cos(\chi+2\rho_k)
+\frac{3}{4}\cos(3\rho_k-\rho_{k0})-\frac{9}{32}\cos(\chi-2\rho_{k0})\nonumber\\
&&+\frac{3}{8}\cos(\chi+3\rho_k-\rho_{k0})-\frac{9}{8}\cos(\chi-3\rho_{k}+\rho_{k0})+\frac{3}{4}\cos(\chi+\rho_{k} +\rho_{k0})+\frac{75}{32}\cos(\chi+2\rho_{k0})\nonumber\\
&&+\frac{21}{16}\cos(2\rho_{k0})+\frac{1}{8}\cos\left(\chi+\rho_k-\rho_{k0}-(k-1)\frac{2\pi}{3}\right)
+\frac{3}{32}\cos\left(\chi+2\rho_k-2\rho_{k0}+(k-1)\frac{2\pi}{3}\right)\nonumber\\
&&-\frac{7}{16}\cos\left(2\rho_k-2\rho_{k0}+(k-1)\frac{2\pi}{3}\right)
-\frac{25}{32}\cos\left(\chi-2\rho_k+2\rho_{k0}-(k-1)\frac{2\pi}{3}\right)\nonumber\\
&&+\cos\left((k-1)\frac{2\pi}{3}\right)\left(\frac{87}{16}+\frac{75}{16}\cos\chi
-5\cos\left(\rho_k-\rho_{k0}\right)-\frac{33}{8}\cos\left(\chi-\rho_k+\rho_{k0}\right)
\right .\nonumber\\
&&\qquad \qquad \qquad \qquad \displaystyle\left.+\frac{1}{4}\cos\left(\rho_k+\rho_{k0}+(k-1)\frac{2\pi}{3}\right)\right)\nonumber\\
&&+(k-1)\cos^k\left(\frac{2\pi}{3}\right)\left(\frac{33\sqrt{3}}{4}\sin\chi
+\frac{\sqrt{3}}{2}\sin\left(\rho_k+\rho_{k0}+(k-1)\frac{2\pi}{3}\right)
+11\sqrt{3}\sin\left(\rho_k-\rho_{k0}\right)\right.\nonumber\\
&&\qquad \qquad \qquad \qquad \left. \displaystyle-\frac{27\sqrt{3}}{4}\sin\left(\chi-\rho_k+\rho_{k0}\right)\right)\nonumber\\
&&+\left[\frac{7}{8}\sin(2\rho_k)+\frac{15}{16}\sin(\chi-2\rho_k)-\frac{27}{16}\sin(\chi+2\rho_k)
-\frac{3}{8}\sin(\chi-\rho_k-\rho_{k0})-3\sin(\rho_k+\rho_{k0})\right.\nonumber\\
&&\qquad -\frac{27}{8}\sin(\chi+\rho_{k}+\rho_{k0})+\frac{3}{8}\sin\left(\chi+\rho_{k} -\rho_{k0}+(k-1)\frac{2\pi}{3}\right)
-3\sin\left(\rho_{k}-\rho_{k0}+(k-1)\frac{2\pi}{3}\right)\nonumber\\
&&\qquad+\frac{27}{8}\sin\left(\chi-\rho_k+\rho_{k0}-(k-1)\frac{2\pi}{3}\right)
+\frac{9}{4}\cos\left((k-1)\frac{2\pi}{3}\right)\sin\chi\nonumber\\
&&\qquad\left.-(k-1)\cos^k\left(\frac{2\pi}{3}\right)\left(\frac{27\sqrt{3}}{4}
+\frac{15\sqrt{3}}{4}\cos\chi\right)\right]\Omega(t-t_0)
\nonumber\\
&&+\left(-\frac{3}{4}\cos(\chi-2\rho_k)+\frac{3}{4}\cos(\chi+2\rho_k)\right)\Omega^2(t-t_0)^2,\label{equ3.34}\\
E_{yk}&=&-\frac{37}{16}\sin(2\rho_k)-\frac{45}{32}\sin(\chi-2\rho_k)-\frac{111}{32}\sin(\chi+2\rho_k)+\frac{3}{4}\sin(3\rho_k-\rho_{k0})
+\frac{9}{32}\sin(\chi-2\rho_{k0})\nonumber\\
&&+\frac{3}{8}\sin(\chi+3\rho_k-\rho_{k0})+\frac{9}{8}\sin(\chi-3\rho_{k}+\rho_{k0})+\frac{3}{4}\sin(\chi+\rho_{k} +\rho_{k0})+\frac{75}{32}\sin(\chi+2\rho_{k0})+\frac{21}{16}\sin(2\rho_{k0})
\nonumber\\
&&-\frac{1}{8}\sin\left(\chi+\rho_k-\rho_{k0}-(k-1)\frac{2\pi}{3}\right)
+\frac{3}{32}\sin\left(\chi+2\rho_k-2\rho_{k0}+(k-1)\frac{2\pi}{3}\right)\nonumber
\end{eqnarray}

\begin{eqnarray}
&&-\frac{7}{16}\sin\left(2\rho_k-2\rho_{k0}+(k-1)\frac{2\pi}{3}\right)
+\frac{25}{32}\sin\left(\chi-2\rho_k+2\rho_{k0}-(k-1)\frac{2\pi}{3}\right)\nonumber\\
&&+(k-1)\cos^k\left(\frac{2\pi}{3}\right)\left(\frac{87\sqrt{3}}{8}+\frac{75\sqrt{3}}{8}\cos\chi
-10\sqrt{3}\cos\left(\rho_k-\rho_{k0}\right)
-\frac{\sqrt{3}}{2}\cos\left(\rho_k+\rho_{k0}+(k-1)\frac{2\pi}{3}\right)\right.\nonumber\\
&&\qquad \qquad \qquad \qquad -\frac{33\sqrt{3}}{4}\cos\left(\chi-\rho_k+\rho_{k0}\right)\bigg) \nonumber\\
&&+\cos\left((k-1)\frac{2\pi}{3}\right)\left(-\frac{33}{8}\sin\chi
+\frac{1}{4}\sin\left(\rho_k+\rho_{k0}+(k-1)\frac{2\pi}{3}\right)
\right.\nonumber\\
&&\qquad \qquad \qquad \qquad\left.-\frac{11}{2}\sin\left(\rho_k-\rho_{k0}\right)+\frac{27}{8}\sin\left(\chi-\rho_k+\rho_{k0}\right)\right)
\nonumber\\
&&+\left[-\frac{7}{8}\cos(2\rho_k)+\frac{15}{16}\cos(\chi-2\rho_k)+\frac{27}{16}\cos(\chi+2\rho_k)
-\frac{3}{8}\cos(\chi-\rho_k-\rho_{k0})+3\cos(\rho_k+\rho_{k0})\right.\nonumber\\
&&\qquad +\frac{27}{8}\cos(\chi+\rho_{k}+\rho_{k0})-\frac{3}{8}\cos\left(\chi+\rho_{k}-\rho_{k0}+(k-1)\frac{2\pi}{3}\right)
+3\cos\left(\rho_{k}-\rho_{k0}+(k-1)\frac{2\pi}{3}\right)\nonumber\\
&&\qquad +\frac{27}{8}\cos\left(\chi-\rho_k+\rho_{k0}-(k-1)\frac{2\pi}{3}\right)+\frac{9\sqrt{3}}{2}(k-1)\cos^k\left(\frac{2\pi}{3}\right)\sin\chi
\nonumber \\
&&\qquad \left.+\cos\left((k-1)\frac{2\pi}{3}\right)\left(\frac{27}{8}+\frac{15}{8}\cos\chi\right)\right]\Omega(t-t_0)
\nonumber\\
\label{equ3.35}&&+\left(\frac{3}{4}\sin(\chi-2\rho_k)+\frac{3}{4}\sin(\chi+2\rho_k)\right)\Omega^2(t-t_0)^2,\\
E_{zk}&=&\tan\phi\left[-\cos\sigma_k+\frac{3}{2}\cos\sigma_{k0}-\frac{3}{2}\cos(\chi-\sigma_k)
+\frac{9}{4}\cos(\chi+\sigma_k)+\frac{3}{4}\cos(\chi-\sigma_{k0})-\frac{1}{2}\cos(2\sigma_k-\sigma_{k0})
\right.\nonumber\\
&&\qquad -\frac{1}{4}\cos(\chi+2\sigma_{k}-\sigma_{k0})+\frac{3}{4}\cos(\chi-2\sigma_{k}+\sigma_{k0})
-\frac{9}{4}\cos(\chi+\sigma_{k0})+\frac{1}{4}\cos\left(\chi-\sigma_k+2\sigma_{k0}\right)
\nonumber\\
&&\qquad \left.+\left(-2\sin\sigma_{k}+\frac{3}{2}\sin(\chi+\sigma_{k})\right)\Omega(t-t_0)
+\left(\frac{3}{4}\cos(\chi-\sigma_k)-\frac{3}{4}\cos(\chi+\sigma_k)\right)\Omega^2(t-t_0)^2\right],\label{equ3.36}\\
E_{xk}'&=&-\frac{87}{32}\cos\varphi-\frac{15}{32}\cos(\varphi-2\chi)-\frac{111}{16}\cos(\varphi-\chi)+\frac{45}{32}\cos(\varphi-2\Omega t)+\frac{5}{32}\cos(\varphi-2\chi-2\Omega t)\nonumber\\
&&+\frac{45}{16}\cos(\varphi-\chi-2\Omega t)-\frac{9}{32}\cos(\varphi-2\Omega t_{0})-\frac{9}{32}\cos(\varphi-2\chi-2\Omega t_{0}) -\frac{45}{16}\cos(\varphi-\chi-2\Omega t_{0})\nonumber\\
&&+\frac{3}{32}\cos(\varphi+2\Omega t-2\Omega t_{0})+\frac{3}{32}\cos(\varphi-2\chi+2\Omega t-2\Omega t_{0})
+\frac{15}{16}\cos\left(\varphi-\chi+2\Omega t-2\Omega t_{0}\right)\nonumber \\
&&-\frac{9}{8}\cos\left(\varphi-\Omega t-\Omega t_{0}\right)
+\frac{1}{8}\cos\left(\varphi-2\chi-\Omega t-\Omega t_{0}\right)
+\frac{21}{8}\cos\left(\varphi+\Omega t-\Omega t_{0}\right)\nonumber\\
&&+\frac{3}{8}\cos\left(\varphi-2\chi+\Omega t-\Omega t_{0}\right)
+6\cos\left(\varphi-\chi+\Omega t+\Omega t_{0}\right)\nonumber \\
&& +\left(\frac{27}{16}\sin\varphi+\frac{3}{16}\sin\left(\varphi-2\chi\right)
+\frac{27}{8}\sin\left(\varphi-\chi\right)-\frac{9}{16}\sin\left(\varphi-2\Omega t\right)-\frac{1}{16}\sin\left(\varphi-2\chi-2\Omega t\right)\right.\nonumber\\
&&\qquad -\frac{9}{8}\sin\left(\varphi-\chi-2\Omega t\right)-\frac{9}{8}\sin\left(\varphi-\Omega t-\Omega t_{0}\right)-\frac{3}{8}\sin(\varphi-2\chi-\Omega t-\Omega t_{0})-\frac{9}{2}\sin\left(\varphi-\chi-\Omega t-\Omega t_{0}\right)\nonumber\\
&&\qquad \left.+\frac{9}{8}\sin\left(\varphi+\Omega t-\Omega t_{0}\right)+\frac{3}{8}\sin\left(\varphi-2\chi+\Omega t-\Omega t_{0}\right)+\frac{9}{2}\sin\left(\varphi-\chi+\Omega t-\Omega t_{0}\right)\right)\Omega (t-t_{0}),
\label{equ3.37}\\
E_{yk}'&=&-\frac{87}{32}\sin\varphi-\frac{15}{32}\sin(\varphi-2\chi)-\frac{111}{16}\sin(\varphi-\chi)
-\frac{45}{32}\sin(\varphi-2\Omega t)-\frac{5}{32}\sin(\varphi-2\chi-2\Omega t)\nonumber\\
&&-\frac{45}{16}\sin(\varphi-\chi-2\Omega t)+\frac{9}{32}\sin(\varphi-2\Omega t_{0})+\frac{9}{32}\sin(\varphi-2\chi-2\Omega t_{0}) +\frac{45}{16}\sin(\varphi-\chi-2\Omega t_{0})\nonumber\\
&&+\frac{3}{32}\sin(\varphi+2\Omega t-2\Omega t_{0})+\frac{3}{32}\sin(\varphi-2\chi+2\Omega t-2\Omega t_{0})
+\frac{15}{16}\sin\left(\varphi-\chi+2\Omega t-2\Omega t_{0}\right)\nonumber\\
&&+\frac{9}{8}\sin\left(\varphi-\Omega t-\Omega t_{0}\right)
-\frac{1}{8}\sin\left(\varphi-2\chi-\Omega t-\Omega t_{0}\right)
+\frac{21}{8}\sin\left(\varphi+\Omega t-\Omega t_{0}\right)\nonumber\\
&&+\frac{3}{8}\sin\left(\varphi-2\chi+\Omega t-\Omega t_{0}\right)
+6\sin\left(\varphi-\chi+\Omega t+\Omega t_{0}\right)\nonumber \\
&&+\left(-\frac{27}{16}\cos\varphi-\frac{3}{16}\cos\left(\varphi-2\chi\right)
-\frac{27}{8}\cos\left(\varphi-\chi\right)-\frac{9}{16}\cos\left(\varphi-2\Omega t\right)-\frac{1}{16}\cos\left(\varphi-2\chi-2\Omega t\right)\right.\nonumber\\
&&\qquad-\frac{9}{8}\cos\left(\varphi-\chi-2\Omega t\right)-\frac{9}{8}\cos\left(\varphi-\Omega t-\Omega t_{0}\right)-\frac{3}{8}\cos(\varphi-2\chi-\Omega t-\Omega t_{0})\nonumber\\
&&\qquad -\frac{9}{2}\cos\left(\varphi-\chi-\Omega t-\Omega t_{0}\right)-\frac{9}{8}\cos\left(\varphi+\Omega t-\Omega t_{0}\right)-\frac{3}{8}\cos\left(\varphi-2\chi+\Omega t-\Omega t_{0}\right)\nonumber\\
&&\qquad \left.-\frac{9}{2}\cos\left(\varphi-\chi+\Omega t-\Omega t_{0}\right)\right)\Omega (t-t_{0}),
\label{equ3.38}\\
\label{equ3.39}E_{zk}'&=&0,
\end{eqnarray}

\begin{eqnarray}
\Lambda_{xk}&=&\frac{3}{2}\cos(\chi-\Omega t)-\frac{5}{2}\cos(\chi+\Omega t)-\frac{3}{4}\cos(\chi-\Omega t_{0})+\frac{1}{4}\cos(\chi+2\Omega t-\Omega t_{0})+\frac{9}{4}\cos(\chi+\Omega t_{0})\nonumber\\
\label{equ3.40}&&-\frac{3}{4}\cos(\chi-2\Omega t+\Omega t_{0})-2\sin(\chi+\Omega t)\Omega (t-t_{0})+\left(-\frac{3}{4}\cos(\chi-\Omega t)+\frac{3}{4}\cos(\chi+\Omega t)\right)\Omega^2 (t-t_{0})^2,\\
\Lambda_{yk}&=&-\frac{3}{2}\sin(\chi-\Omega t)-\frac{5}{2}\sin(\chi+\Omega t)+\frac{3}{4}\sin(\chi-\Omega t_{0})+\frac{1}{4}\sin(\chi+2\Omega t-\Omega t_{0})+\frac{9}{4}\sin(\chi+\Omega t_{0})\nonumber\\
\label{equ3.41}&&+\frac{3}{4}\sin(\chi-2\Omega t+\Omega t_{0})+2\cos(\chi+\Omega t)\Omega (t-t_{0})+\left(\frac{3}{4}\sin(\chi-\Omega t)+\frac{3}{4}\sin(\chi+\Omega t)\right)\Omega^2 (t-t_{0})^2,\\
\label{equ3.42}\Lambda_{zk}&=&0.
\end{eqnarray}
Here, the parameter angles $\rho_{k},\rho_{k0}$ and $\sigma_{k0}$ are defined as
\begin{equation}\label{equ3.43}
\rho_{k}=\Omega t+(k-1)\frac{2\pi}{3},\qquad \rho_{k0}=\Omega t_{0}+(k-1)\frac{2\pi}{3},\qquad \sigma_{k0}=\Omega t_{0}-(k-1)\frac{2\pi}{3},\quad \text{for}\ k=1,2,3.
\end{equation}
\end{widetext}

\end{document}